\documentclass[conference]{IEEEtran}
\IEEEoverridecommandlockouts
\usepackage{cite}
\usepackage{amsmath,amssymb,amsfonts}
\usepackage{algorithmic}
\usepackage{graphicx}
\usepackage{textcomp}
\usepackage{xcolor}
\ifCLASSOPTIONcompsoc
	\usepackage[caption=false,font=normalsize,labelfont=sf,textfont=sf]{subfig}
\else
	\usepackage[caption=false,font=footnotesize]{subfig}
\fi
\def\BibTeX{{\rm B\kern-.05em{\sc i\kern-.025em b}\kern-.08em
    T\kern-.1667em\lower.7ex\hbox{E}\kern-.125emX}}
\begin{document}

%\title{Flattened Exponential Histogram for Sliding Window Queries over Data Streams
%}
\title{Flattened Exponential Histogram for Sliding Window Queries over Data Streams\\
	\thanks{Corresponding author: Dagang Li.}
}
\author{\IEEEauthorblockN{Shuhao Sun\IEEEauthorrefmark{1}, Dagang Li\IEEEauthorrefmark{2}\IEEEauthorrefmark{1}}
	\IEEEauthorblockA{\textit{\IEEEauthorrefmark{1}School of Electronic and Computer Engineering, Peking University} \\
		Shenzhen, China \\
		\textit{\IEEEauthorrefmark{2}Macau University of Science and Technology}\\
		Macau SAR, China \\
		shuhaosun@pku.edu.cn, dagang.li@ieee.org}
	}
\maketitle

\begin{abstract}
The Basic Counting problem [1] is one of the most fundamental and critical streaming problems of sliding window queries over data streams. Given a stream of 0's and 1's, the purpose of this problem is to estimate the number of 1's in the last N elements (or time units) seen from the stream. Its solution can be used as building blocks to solve numerous more complex problems such as heavy hitter, frequency estimation, distinct counting, etc. In this paper, we present the flattened exponential histogram (FEH) model for the Basic Counting problem. Our model improves over the exponential histogram [1], [2], a well-received deterministic technique for Basic Counting problem, with respect to accuracy and memory utilization most of the time in practice. Extensive experimental results on real-world datasets show that with the same memory footprint, the accuracy of our model is between 4 to 15 and on average 7 times better than that of the exponential histogram, while the speed is roughly the same.
\end{abstract}


\section{Introduction}

\subsection{Background and Motivation}
Data-stream processing has been a hot research field for years. Many effective algorithms and data structures have been proposed for estimating aggregates and statistics over data streams in many fields, such as databases [3]-[5], data mining [6]-[8], network measurement [9]-[11], etc. Because of the large scale and very high speed of streams, the space and the time overhead of storing the entire stream to compute the exact statistics are huge. Therefore, these algorithms and structures typically provide approximate estimation with guarantees on the accuracy of the query answer, using small space and processing streams in real-time.

Most data-stream models maintain statistics of all data elements seen so far, and consider them equally important. However, as more and more data elements arrive, the counting model will eventually run out of capacity or become very inaccurate. On the other hand, in many practical scenarios, recent data is actually more important than old data. For instance, in the stock-trading system, people tend to be more interested in recent trading statistics rather than the historical ones. Time-decay models have been proposed in the literature to deal with these problems. The weight of data elements in statistics decreases with age by polynomial decay or exponential decay [12]-[15]. The sliding-window model is one of the widely used time-decay model, which keeps the statistics of only the last N elements in the stream (a count-based window) or the elements that have arrived in the last N time units (a time-based window), and all of which are of the same weight. Extensive sliding windows models have been proposed for various streaming problems [16]-[21]. Similar to other approximate algorithms and structures in data stream fields, the sliding window model attempts to achieve high accuracy, small memory usage and real-time processing.

One of the most fundamental and critical streaming problems of sliding window queries over data streams is the Basic Counting problem [1]. Given a stream of 0's and 1's, the problem is to estimate the number of 1's in the last N elements/time units seen from the stream. Its solution can be used as a building block to solve numerous more complex problems such as top-k [22]-[24], heavy hitter [25], [26], frequency estimation [27]-[29], etc. Datar et al. [1] proposed the first model for Basic Counting problem, the exponential histogram. It provides a $(1+\epsilon)$-multiplicative approximation, which means that the relative error of the estimation returned by the query will never exceed $\epsilon$. For approximate algorithms, the relative error is an important metric to measure the accuracy of statistics. In general, it can better reflect the credibility of the estimation. Using the $(1+\epsilon)$-multiplicative approximation algorithm, we can set the value of $\epsilon$ according to the practical application requirements, so that the relative error can be limited by the required upper bound. There are many other $(1+\epsilon)$-multiplicative approximation algorithms that can be used for Basic Counting problem [30], [31]. The exponential histogram is a state-of-the-art deterministic technique for $(1+\epsilon)$-multiplicative approximation of Basic Counting problem, which is extensively used in data stream processing [32]-[34].

Through in-depth study, we found that the exponential histogram faces the following problems. In practical applications, the measurement system always sets parameters and allocates system resources in advance according to experience or prediction of the measurement. For instance, the dynamic memory allocation and pointers are difficult to implement in network devices, and hardware implementations often adopt static memory allocation [26]. Therefore, in applications, the exponential histogram pre-estimates the possible statistical maximum and allocates memory accordingly. However, during the processing of the exponential histogram, for most of the time, the actual count is much smaller and the memory usage is smaller or even significantly smaller than the allocated memory, resulting in low memory utilization. In addition, we also found that there is still room for improvement in the upper bound of absolute error given by the exponential histogram. 

\subsection{Our Solution}
In this paper, we propose the Flattened Exponential Histogram (FEH), a count-based sliding window model that handles the above mentioned problems of the exponential histogram (EH) model. 

First, the FEH applies an update strategy of “flattened count”, which makes use of the memory space that is otherwise not used in EH. Under different data stream distributions, the space of FEH can always be used to the maximum extent, so that the statistics are more fine-grained, thus improving the accuracy. Second, the FEH adopts an optimized query strategy, which minimizes the absolute error of the query answer and further improves the statistical accuracy. Third, we provide practical solutions to FEH implementation, using the ideas of a cyclic array and word acceleration to optimize the memory operation and speed of FEH. 

We conduct extensive experiments on two real datasets and one synthetic dataset. The experimental results show that our flattened exponential histogram model achieves between 4 to 15 and on average 7 times higher accuracy than the EH model when using the same memory, while the speed is generally not affected. Our model can also be used as building blocks to solve numerous more complex problems in the field of data streams.

\section{RELATED WORK}
Many studies directly address the Basic Counting problem. Exponential histogram (EH) [1], [2] is the first solution for this problem, which guarantees $(1+\epsilon)$-multiplicative approximation. If N represents the window size, it uses $O\left(\frac{1}{\epsilon} \log ^{2} N \epsilon\right)$ bits of space. The EH works with $O(1)$ amortized update time and $O(\log N)$ worst case update time, and can generate query answer over the current window in $O(1)$ time. Waves [30] are a novel family of synopsis data structures to solve the Basic Counting problem, which can also achieve $(1+\epsilon)$-multiplicative approximation. They improve the worst-case update time to a constant, and the complexity of other performance is the same as that of the EH. However, Papapetrou et al. [33] found that EH is faster and more compact in practical applications, requiring about half of the space compared with waves. Lee and Ting [31] proposed an algorithm with smaller space, but it can only achieve $(1+\epsilon)$-multiplicative approximation if the number of 1's in the window is remarkable. Basat et al. [35] provided a $N\epsilon$-Additive Approximation algorithm for Basic Counting problem. Let $f$ be the true value of statistics and N be the window size, then the $N\epsilon$-Additive Approximation means that the estimation $\hat{f}$ satisfies $|f-\hat{f}|<\epsilon$. The algorithm works with $O(1)$ amortized update time and $O(\log N)$ worst case update time, and uses $O\left(\frac{1}{\epsilon}+\log N \epsilon\right)$ bits of space.

In addition, there are many studies not directly address the Basic Counting problem, but their solutions can be used to cope with this problem after some modifications. For instance, in [26], two novel algorithms proposed for finding heavy hitters and their estimated frequencies over sliding windows. These algorithms guarantee $(N, \epsilon)$-approximation, which means that their estimated frequencies $\hat{f}$ satisfies $f \leq \hat{f} \leq f+N \epsilon$. If the algorithms are modified to only estimate the frequency of one specified element, then it is equivalent to solving the Basic Counting problem. Assaf et al. [36] propose a composite structure that solves many common problems over sliding windows, such as membership query, frequency counting, distinct counting, etc. This structure also guarantees $(N, \epsilon)$-approximation for frequency counting, and can be adapted to the Basic Counting problem. Zhou et al. [37] present two novel solutions for the per-flow counting problem, which achieve high accuracy and fast processing speed using limited memory. They can be modified to deal with the Basic Counting problem. Persistent sketches [38] were designed to query statistics over any period of time. If the sketches only query the statistics of the last window and only count the frequency of one element, then it solves the Basic Counting problem. There are other studies that can be occupied to this problem [23], [39]-[42], but they are not used to solve this problem naturally and are basically not guarantee $(1+\epsilon)$-multiplicative approximation.

In this paper, we only focus on the Basic Counting $(1+\epsilon)$-multiplicative approximation, which is directly connected to the vital metric of relative error. Among the related algorithms, EH has the best performance.

\section{PRELIMINARIES}
\subsection{Basic Counting Problem}
Given a stream of 0's and 1's, the basic counting problem is to count the number of 1's in the sliding window. As shown in Fig.1, data elements arrive from the right, which means that the elements on the right are newer and the left are older. This is a count-based sliding window with a size of 30 elements. Arrival time represents the ordinal number of each element in the window, which increases with the arrival of elements. When the arrival time reaches the window size, it will circle back to 1 and then increase again. In Fig.1, the current arrival element is 0 with arrival time 1, and the next arrival element is 1 with arrival time 2. 

\begin{figure}[htbp]
	\centerline{\includegraphics[width=\linewidth]{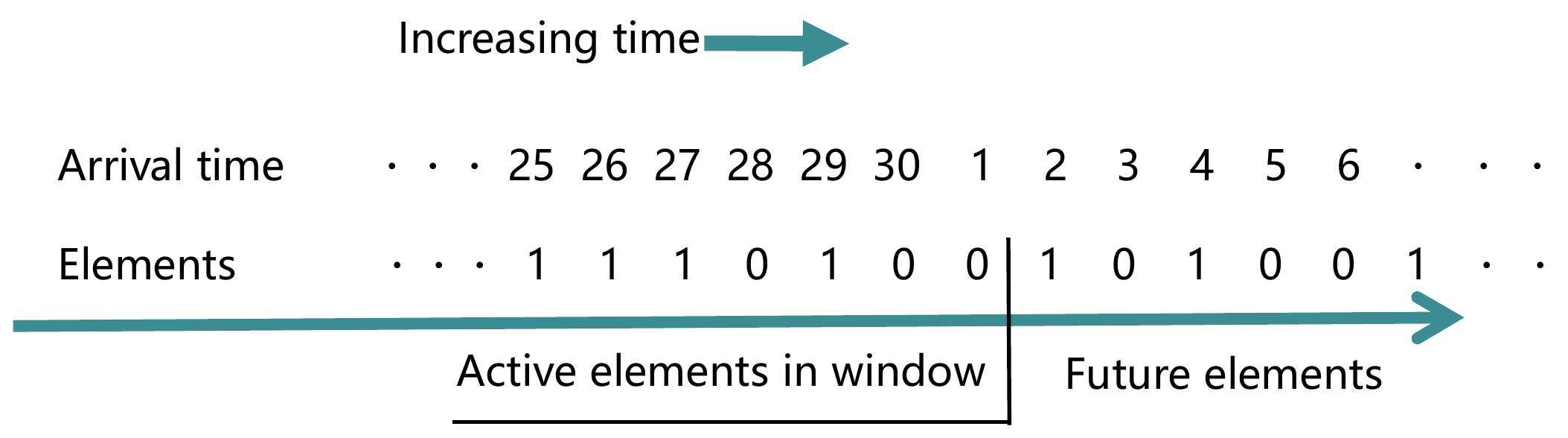}}
	\caption{An illustration for the Basic Counting problem}
	\label{fig1}
\end{figure}

\subsection{Exponential Histogram}
The structure of the EH model can be considered as a bucket array, where each bucket contains two records, bucket size and timestamp. The bucket size represents the number of active 1's in that bucket, and the timestamp corresponds to the arrival time of the most recent 1 in that bucket. By comparing with the arrival time of the current element, it can be determined whether the bucket timestamp is out of the window (expired) and if that happens, that means there is no active 1 in it, and the whole bucket can be removed. In EH, at any moment, there is at most one bucket, that is, the oldest bucket, that may contain some 1's that have expired. EH maintains two counters for queries. One is used to record the size of the oldest bucket (Last), and the other is used to record the sum of the sizes of all buckets (Total). EH returns $Total-Last/2$ as the query result of the basic counting problem. Therefore, the query complexity of EH is $O(1)$, and the maximum absolute error of EH is $Last/2$. If the bucket array is numbered, for instance, the newest bucket size is numbered $C_1$ and the oldest bucket size is $C_j$, then the upper bound of relative error of EH at any instant is expressed as by the following formula:
\begin{equation}
relative \ error<=\frac{C_{j}}{2\left(1+\sum_{i=1}^{j-1} C_{i}\right)}  \label{eq1}
\end{equation}

EH defines $k=\left\lceil\frac{1}{\epsilon}\right\rceil$, thus the $(1+\epsilon)$-multiplicative approximation is converted to making the relative error less than or equal to $\frac{1}{k}$ at any time. In order to achieve that, EH makes the bucket size satisfy $C_{1} \leq C_{2} \leq \ldots \leq C_{j}$ and $C_{i} \in\left\{1,2,4, \ldots 2^{m}\right\}$. For each bucket size other than the size of the oldest and newest bucket, there are at least $\frac{k}{2}$ and at most $\frac{k}{2}+1$ buckets of that size(if $\frac{k}{2}$ is not an integer, replace $\frac{k}{2}$ by $\left\lceil\frac{k}{2}\right\rceil$); for the size of the newest bucket (equal to 1), there are at least $k$ and at most $k+1$ buckets; for the size of the oldest bucket, there are at most $\frac{k}{2}+1$ buckets of that size. If the window size is $N$ and $C_{i} \in\left\{1,2,4, \ldots, 2^{m^{\prime}}\right\}$, then $m^{\prime} \leq \log \frac{2 N}{k}+1$ can be obtained. From all the above restrictions, the following formula can be deduced:
\begin{equation}
\frac{C_{j}}{2\left(1+\sum_{i=1}^{j-1} C_{i}\right)} \leq \frac{1}{k} \label{eq2}
\end{equation}
Hence, the EH achieves $(1+\epsilon)$-multiplicative approximation. 

The number of buckets required for EH model is:
\begin{equation}
	m=\left(\frac{k}{2}+1\right) \cdot\left(\log \left(\frac{2 N^{\prime}}{k}\right)+2\right)
\end{equation}
$N^{\prime}$ is the maximum number of possible occurrences in the window. Each bucket keeps $O(loglogN)$ bucket size and $O(logN)$ timestamp. Consequently, the overall space requirement of the EH model is $O\left(\frac{1}{\epsilon} \log ^{2} N \epsilon\right)$ bits. The following is an example of a EH insertion process. The bucket size from right to left is $C_{1}, C_{2}, \ldots, C_{j}$ and $k=2$. Each step represents when a new element 1 arrives, the insertion process of EH.
\begin{itemize}
	\item \emph{Step 1}:	\	16 8 8 4 4 2 1 1 1
	\item \emph{Step 2}:	\	16 8 8 4 4 2 1 1 1 1 	
	
	The number of 1's is more than $k+1$ and needs to be merged:
	\item \emph{Step 3}:	\	16 8 8 4 4 2 2 1 1
	\item \emph{Step 4}:	\	16 8 8 4 4 2 2 1 1 1
	\item \emph{Step 5}:	\	16 8 8 4 4 2 2 1 1 1 1
	
	The number of 1's is more than $k+1$ and needs to be merged:
	\item \emph{Step 6}:	\	16 8 8 4 4 2 2 2 1 1
	
	The number of 2's is more than $\frac{k}{2}+1$ and needs to be merged:
	\item \emph{Step 7}:	\	16 8 8 4 4 4 2 1 1
	
	The number of 4's is more than $\frac{k}{2}+1$ and needs to be merged:
	\item \emph{Step 8}:	\	16 8 8 8 4 2 1 1
	
	The number of 8's is more than $\frac{k}{2}+1$ and needs to be merged:
	\item \emph{Step 9}:	\	16 16 8 4 2 1 1
\end{itemize}

When the number of bucket sizes exceeds the corresponding maximum, the merge operation is performed. The size of the merged bucket is equal to the sum of the original two bucket sizes, and the timestamp of the merged bucket is the timestamp of the newer bucket (the right bucket). One insertion may result in a series of merge operations; such as step 5 to 8. The EH works with $O(\log N)$ worst case update time, and the amortized update time is O(1).

\section{FLATTENED EXPONENTIAL HISTOGRAM}
In this section, the flattened exponential histogram will be described from two perspectives, the update and the query strategy. Furthermore, a practical design of the FEH will be provided.

\subsection{Main Idea}
The arrangement on the size of the buckets in EH guarantees that the denominator of (1) is always at least $k$ times larger than $C_j$ in the numerator. Indeed as shown in Fig.2(a), if we call the buckets of the same size as a row, then the larger the $k$, the longer each row can be, so the shaded area (representing the total count, a.k.a. the denominator) to the right of the oldest bucket is larger and the relative error is smaller. Only when the oldest bucket is exactly at the right edge of each row, will we get to the maximal error of $\frac{1}{k}$, for all the other count value the relative error is always smaller than $\frac{1}{k}$, the further away the oldest bucket from the right edge the better. 

The design of EH is perfect and sound to achieve the guarantee of $\frac{1}{k}$ maximal relative error, but we found the empty buckets to the left of the oldest bucket a waste: instead of pushing to the right towards the error upper bound, if we redistribute the count value as pouring water into the triangular space below the upper bound as shown in Fig.2(b), for the same size of the shaded area, we may get a smaller $C_j$ (a.k.a. the numerator) by making full use of all the pre-allocated buckets. Although this change will not improve the maximal error guarantee (at the row edge), it is still easy to see that the relative error for all other count values will improve, as the new pouring-water fashion will withhold the growth of $C_j$ as much as possible when the total count accumulates. Only when a full layer of $C_j$ is filled up, will we start to double $C_j$ and pave another layer as shown in Fig.2(c), starting from the left. By doing so we effectively hold the right edge of the highest row furthest away from the error upper bound, so when old buckets expire one by one, $C_j$ will monotonously drop down at the earliest possible chance.

Comparing to EH, the pouring-water fashion can be seen as flattening the height of EH across all buckets so we call it Flattened Exponential Histogram, or FEH. Now that we can't leave empty buckets to the left when old buckets expire, the update mechanism needs to be redesigned. Furthermore, handling rows of elastic length in FEH also bring challenges to the practical efficiency, and in the following subsections we will explain how to handle these issues.

\begin{figure}[tbp]
	\centerline{\includegraphics[width=\linewidth]{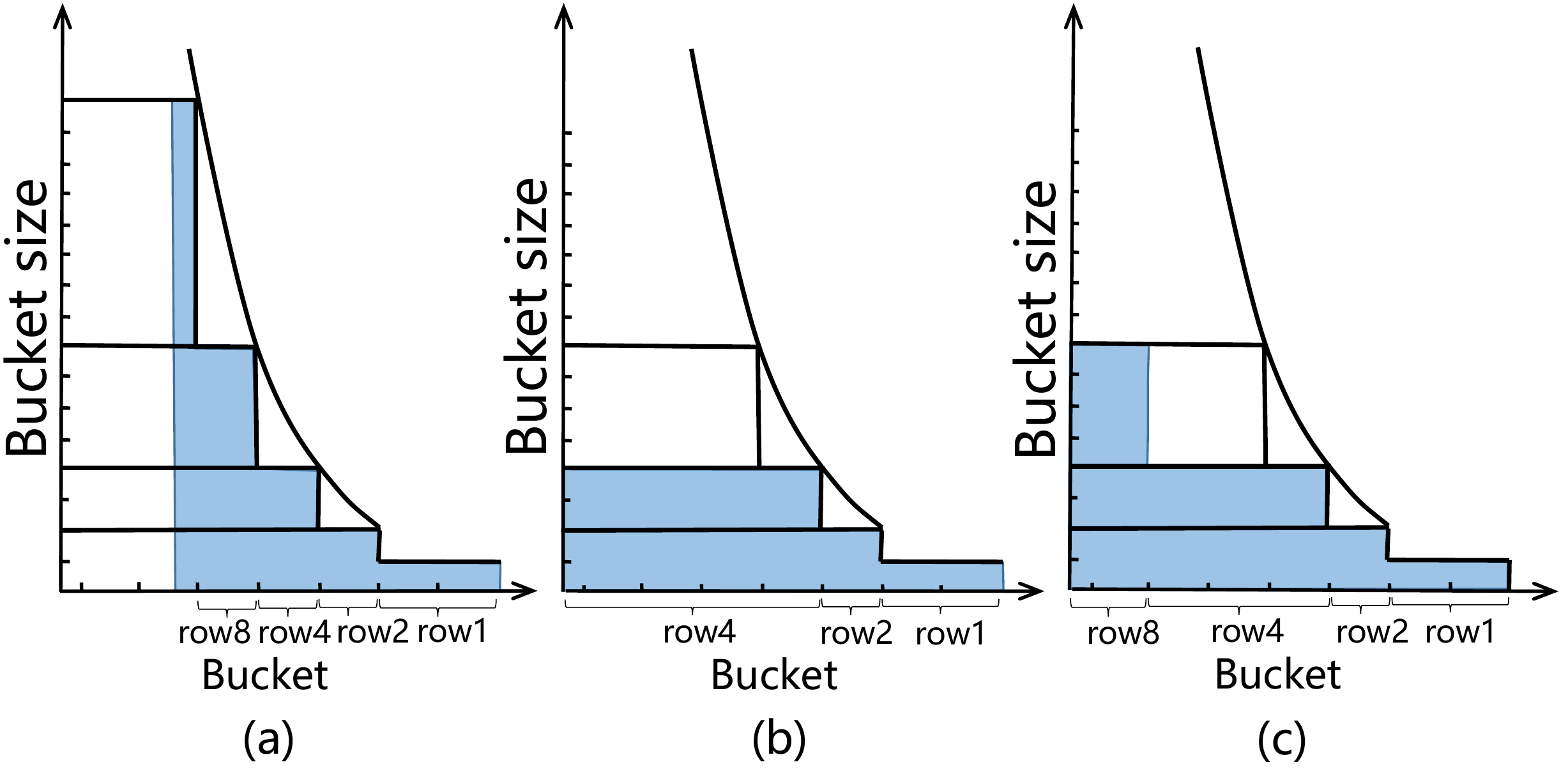}}
	\caption{The basic difference between the idea of EH and FEH}
	\label{fig2}
\end{figure}

\subsection{The Update Strategy of FEH}
In applications, the maximum number of 1's that may appear in the window, is estimated in advance, and the memory is allocated for the EH model according to (3). A large amount of memory on the left side of the bucket array is not utilized when the total count is small. From this point of view, we hope to make full use of memory at any time to reduce the size of the oldest bucket, thus reducing errors and improving accuracy. The key idea of the update strategy of FEH is that, if empty buckets exist, they should be used during updates to store the arrival of new 1’s; when the number of empty buckets is insufficient, FEH merge operations are performed to generate empty buckets by merging adjacent buckets. We call this ``flattened count''. Next, we will describe the flattened count step by step in two cases: arrival and expiration.
\paragraph{Update strategy in sample cases} When the bucket expiration is not considered, we call it a simple case. In this case, when new element 1 arrives, the update rule for FEH is described below.

\begin{itemize}
	\item Before the bucket array is first fully filled with size=1 and their timestamps, fill the bucket array with size=1 and their timestamps from left to right.
	\item After the bucket array is first fully filled with size=1 and their timestamps, first follow the EH update mode to update from right to left. If the number of empty buckets is enough, the update is completed. Otherwise, the empty bucket will be generated by the FEH merge operation, and then continue the EH update mode to complete the remaining update steps.
\end{itemize}

When the first merge operation is performed, bucket array has been completely filled with size=1 and their timestamps. For the direction of subsequent updates, we make the bucket size on the right side of the bucket array the same as the size in EH (the bold part in the following examples, referred to as ``the EH part''), and the largest size in this part is ``partition size''. The left side of the bucket array (referred to as ``the FEH part'') is filled with this partition size. For example, when $\frac{k}{2}+1$, $ \log \left(\frac{2 N^{\prime}}{k}\right)+2=7$:

\begin{itemize}
\item 1111 \ \ 1111 \ \ 1111 \ \ 1111 \ \  1111 \ \ \textbf{1111 \ \ 1111} 

(bucket array fully filled up) \ update direction $-->$ 

\item 2222 \ \ 2222 \ \ 2222 \ \ 2222  \ \ \textbf{2222 \ \ 1111 \ \ 1111} 

(partition size=2) \ update direction $-->$

\item 4444 \ \ 4444 \ \ 4444 \ \ \textbf{4444 \ \ 2222 \ \ 1111 \ \ 1111}  

(partition size=4) \ update direction $-->$

\item 8888 \ \ 8888 \ \ \textbf{8888 \ \ 4444 \ \ 2222 \ \ 1111 \ \ 1111}  

(partition size=8)
\end{itemize}

The rest may be deduced by analogy. When the number of 1's in the window reaches the upper limit N', the bucket array of the FEH model and the EH model will become the same. Updating in this direction, the memory of the FEH part that was not used in the EH model was fully utilized in the FEH model. Therefore, before the total count reached N', the absolute error of the FEH is significantly lower than that of the EH, thus improving accuracy. In order to achieve the above update result, the FEH merge operation is the crucial step, which merge the left-most buckets (the oldest two buckets) of the partition size and place the empty bucket generated to the right of the right-most bucket (the newest bucket) of the partition size. Here are two examples.
\begin{enumerate}

	 \item Example 1:
	
     \emph{Step 1}: \ 2222 \ \ 2222 \ \ 2222 \ \ 2222 \ \ \textbf{2222 \ \ 1111 \ \ 1111} 
	
	Now, here comes a new element 1. First, the EH part follow the EH update mode to update. When the merged bucket size in the EH update mode is as large as the partition size, if there has an empty bucket of this size, the update will be completed. Otherwise, carry out the FEH merge operation. In this example, partition size=2, and there are no empty buckets of this size, thus performing the FEH merging operation:
	
	 4222 \ \ 2222 \ \ 2222 \ \ 2222 \ \ \textbf{2220 \ \ 1110 \ \ 1111}  
	
	 \textbf{2\ (Unwritten bucket)}
	
	Partition size=2. The bucket of size 0 on the right side of the newest bucket of size 2 is the empty bucket after FEH merging. The unwritten bucket is generated by the unfinished EH update mode. Then, continue the EH update mode, writing the unwritten bucket to that empty bucket and this update is completed:
	
	\emph{Step 2}: \ 4222 \ \ 2222 \ \ 2222 \ \ 2222 \ \ \textbf{2222 \ \ 1110 \ \ 1111}
	
	Now, here comes a new element 1. Using the EH update mode to update, the number of empty buckets is sufficient to complete the update, thus there is no need for FEH merge:
	
	\emph{Step 3}: \ 4222 \ \ 2222 \ \ 2222 \ \ 2222 \ \ \textbf{2222 \ \ 1111 \ \ 1111}
	
	\item Example 2:
	
	\emph{Step 1}: \ 4444 \ \ 4444 \ \ 4444 \ \ \textbf{4444 \ \ 2222 \ \ 1111 \ \ 1111} 
	
	Now, here comes a new element 1. There are no empty buckets of size 4, thus performing the FEH merging operation:
	
	8444 \ \ 4444 \ \ 4444 \ \ \textbf{4440 \ \ 2220 \ \ 1110 \ \ 1111}
	
	\textbf{4\ (Unwritten bucket)}
	
	Partition size=4. The bucket of size 0 on the right side of the newest bucket of size 4 is the empty bucket after FEH merging. Then, continue the EH update mode, writing the unwritten bucket to that empty bucket and this update is completed:
	
	\emph{Step 2}: \ 8444 \ \ 4444 \ \ 4444 \ \ \textbf{4444 \ \ 2220 \ \ 1110 \ \ 1111}
	
	Now, here comes a new element 1. Using the EH update mode to update, the number of empty buckets is sufficient to complete the update, thus there is no need for FEH merge:
	
	\emph{Step 3}: \ 8444 \ \ 4444 \ \ 4444 \ \ \textbf{4444 \ \ 2220 \ \ 1111 \ \ 1111}  
	
	Now, here comes a new element 1:
	
	\emph{Step 4}: \ 8444 \ \ 4444 \ \ 4444 \ \ \textbf{4444 \ \ 2222 \ \ 1110 \ \ 1111}   
	
	Now, here comes a new element 1:
	
	\emph{Step 5}: \ 8444 \ \ 4444 \ \ 4444 \ \ \textbf{4444 \ \ 2222 \ \ 1111 \ \ 1111}	
\end{enumerate}
	
\paragraph{Update strategy in practical cases} When the bucket expiration is considered, it is the practical case. In practice, the ultimately update rule for FEH is described below.

\begin{itemize}
	\item Check the timestamp of the left-most bucket in the bucket array, empty it if it expires, and move it to the right-most side of its original bucket size.
	\item If the element is 0, ignore it; otherwise, follow the update rule in the simple cases.
\end{itemize}	

For example, when $\frac{k}{2}+1 = 4$, $\log \left(\frac{2 N^{\prime}}{k}\right)+2=7$:

%\begin{itemize}
   \emph{Step 1}: 8888 \ \ 8888 \ \ \textbf{8888 \ \ 4444 \ \ 2222 \ \ 1111 \ \ 1111}

Now, here comes a new element 1. The oldest bucket was expired, and move it to the right-most side of the bucket with size 8:

  8888 \ \ 8888 \ \ \textbf{8880 \ \ 4444 \ \ 2222 \ \ 1111 \ \ 1111}

Then follow the update rule in the simple cases. The EH updates alone are sufficient to complete the updates without the need for FEH merge operations:

   \emph{Step 2}: 8888 \ \ 8888 \ \ \textbf{8880 \ \ 4440 \ \ 2220 \ \ 1110 \ \ 1111}

%\end{itemize}

The memory usage of FEH model is the same as that of EH model, which is $O\left(\frac{1}{\epsilon} \log ^{2} N \epsilon\right)$. After the bucket array is first fully filled with size=1 and their timestamps, the update operation of the FEH model can be divided into the EH update and the FEH merge operation. The complexity of the FEH operation is $O(1)$, because for one update, only one merge is needed to generate empty bucket for subsequent updates. Therefore, the complexity of the FEH update is the same as that of the EH model, in the worst case is $O(logN)$ and on average is $O(1)$. In terms of accuracy, the memory of the FEH part is fully utilized in the FEH model. The error of the FEH model is much smaller than the EH model in most case and the worst case is the same as the EH model.

\subsection{The Query Strategy of FEH}
The query strategy of the EH model is very simple. The $total =\frac{C_{j}}{2}+\sum_{i=1}^{j-1} C_{i}$ will be returned for each query, and the upper bound of the absolute error is $\frac{C_{j}}{2}$. However, we found that this query strategy does not always minimize the upper bound of absolute error and is not optimal. We design a new query strategy in the FEH model to minimize the upper bound of absolute error at any time, which will further improve the accuracy.

In FEH, whenever an expired bucket size is cleared, its timestamp $t1$ will be recorded. Suppose $T_{j}$ is the timestamp of the current oldest bucket. If $T_{j}>t1$, let $t2=T_{j}$, otherwise. let $t2=T_{j}+window \ size$. Then, the growth of the current oldest bucket size can be limited into range $(t1, t2]$. When $C_{j} \leq \frac{1}{2}\left(t_{2}-t_{1}\right)$ , the growth line is limited to the parallelogram of Fig. 3(a). When $C_{j}>\frac{1}{2}\left(t_{2}-t_{1}\right)$ , the growth line is limited to the parallelogram of Fig. 3(b). When being queried, the left boundary of the current window in the range of $(t1, t2]$ will be $x$, and the active 1's in the window can be calculated by $C_{j}-f(x)$. Therefore, in order to minimize the absolute error, we use $f^{\prime}(x)$ in Fig.3 as an estimate of $f(x)$, and it can be expressed as $f^{\prime}(x)=\frac{f_{1}(x)+f_{2}(x)}{2}$.

\begin{figure}[!t]
	\centering
	\subfloat[]{\includegraphics[width=1.5in]{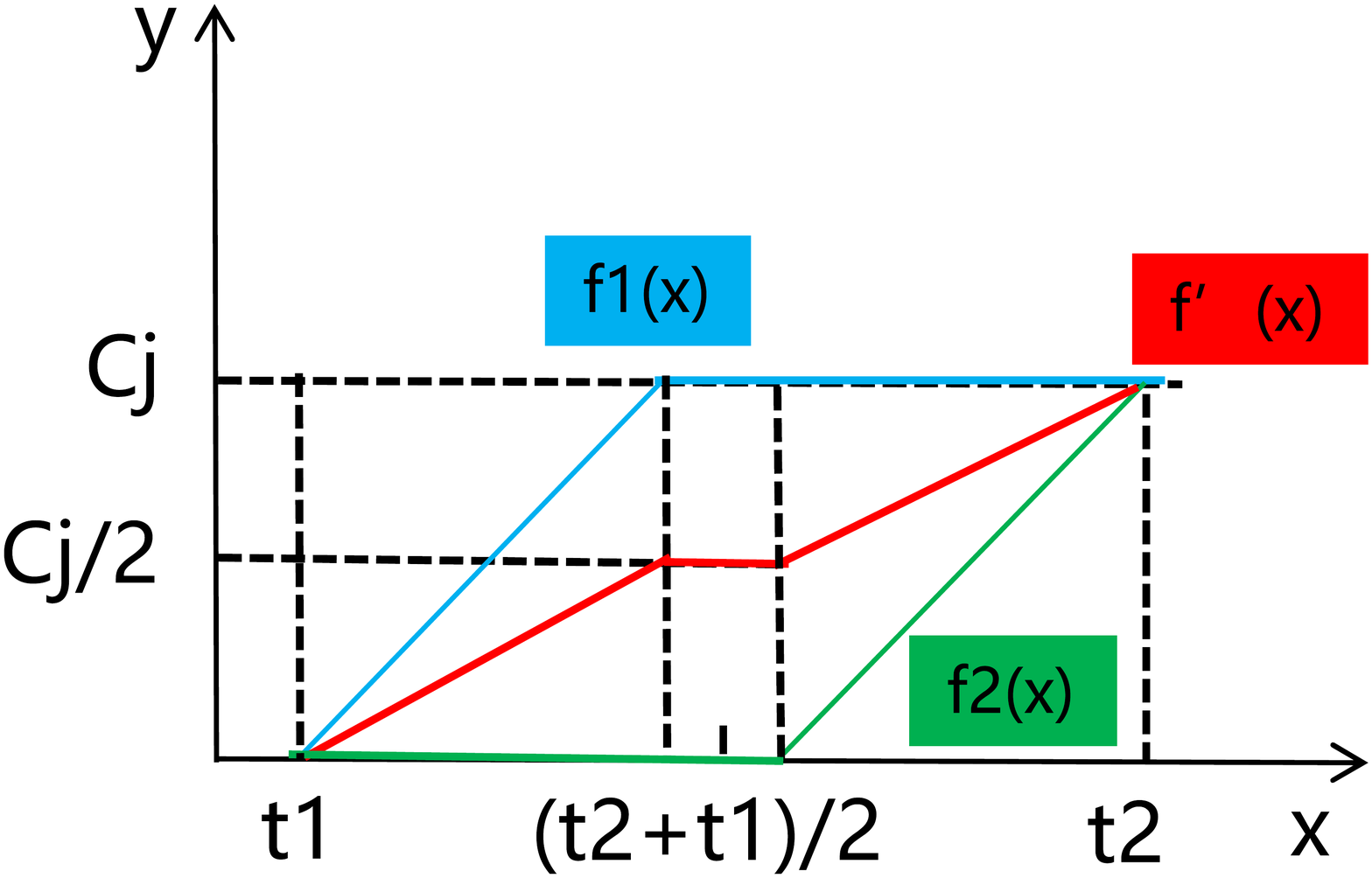}
		\label{fig3_1}}
%	\hfil
	\subfloat[]{\includegraphics[width=1.5in]{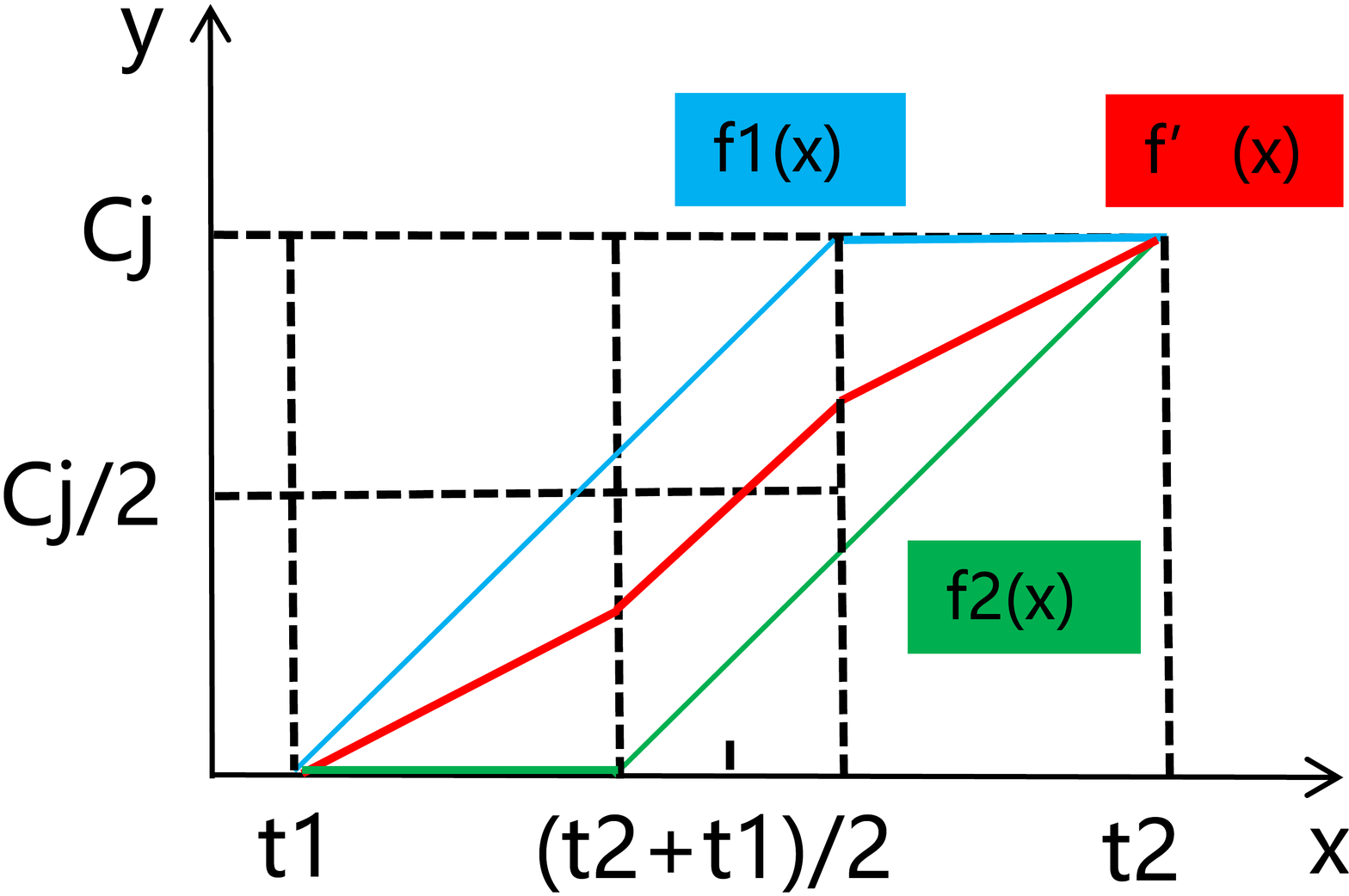}
		\label{fig3_2}}
	\caption{The Growth Line of the Oldest Bucket Size}
	\label{fig3}
\end{figure}

When $C_{j} \leq \frac{1}{2}\left(t_{2}-t_{1}\right)$:
$$
f(x)=\left\{\begin{array}{ll}{\frac{x-t_{1}}{2}} & {\left[t_{1}, t_{1}+C_{j}\right)} \\ {\frac{C_{j}}{2}} & {\left[t_{1}+C_{j}, t_{2}-C_{j}\right]} \\ {\frac{1}{2}\left(x-t_{2}\right)+C_{j}} & {\left(t_{2}-C_{j}, t_{2}\right)}\end{array}\right.
$$

When $C_{j}>\frac{1}{2}\left(t_{2}-t_{1}\right)$:
$$
f(x)=\left\{\begin{array}{ll}{\frac{x-t_{1}}{2}} & {\left[t_{1}, t_{2}-C_{j}\right)} \\ {x-\frac{\left(t_{1}+t_{2}\right)-C_{j}}{2}} & {\left[t_{2}-C_{j}, t_{1}+C_{j}\right]} \\ {\frac{1}{2}\left(x-t_{2}\right)+C_{j}} & {\left(t_{1}+C_{j}, t_{2}\right)}\end{array}\right.
$$

FEH still uses the Total and Last counter to maintain the size of the oldest bucket and the total count in the window. When being queried, it first calculates the current left boundary $x$ of the window. If $T_{j}>t1$, $x=current\ timestamp$, otherwise, $x=current \ timestamp+window \ size$. The FEH returns the final query answer as $Total-f^{\prime}(x)$. The upper bound of the absolute error $AE(x)$ for the query answer is the maximum deviation between $f^{\prime}(x)$ and $f(x)$. According to Fig.3, it can be obtained:

When $C_{j} \leq \frac{1}{2}\left(t_{2}-t_{1}\right)$:
$$
A E(x)=\left\{\begin{array}{ll}{\frac{x-t_{1}}{2}} & {\left[t_{1}, t_{1}+C_{j}\right)} \\ {\frac{C_{j}}{2}} & {\left[t_{1}+C_{j}, t_{2}-C_{j}\right]} \\ {\frac{t_{2}-x}{2}} & {\left(t_{2}-C_{j}, t_{2}\right)}\end{array}\right.
$$

When $C_{j}>\frac{1}{2}\left(t_{2}-t_{1}\right)$:
$$
A E(x)=\left\{\begin{array}{ll}{\frac{x-t_{1}}{2}} & {\left[t_{1}, t_{2}-C_{j}\right)} \\ {\frac{t_{2}-t_{1}-C_{j}}{2}} & {\left[t_{2}-C_{j}, t_{1}+C_{j}\right]} \\ {\frac{t_{2}-x}{2}} & {\left(t_{1}+C_{j}, t_{2}\right)}\end{array}\right.
$$

Fig. 4(a) shows the upper bound for absolute error within $(t 1, t 2]$ when $C_{j} \leq \frac{1}{2}\left(t_{2}-t_{1}\right)$, and Fig. 4(b) shows the upper bound for absolute error within $(t 1, t 2]$ when $C_{j}>\frac{1}{2}\left(t_{2}-t_{1}\right)$. The upper bound of the absolute error is the same as EH model only in case of $C_{j} \leq \frac{1}{2}\left(t_{2}-t_{1}\right)$ and $x \in\left[t_{1}+C_{j}, t_{2}-C_{j}\right]$, which is $\frac{C_{j}}{2}$. In most cases, the upper bound of absolute error is less than $\frac{C_{j}}{2}$.

\begin{figure}[!t]
	\centering
	\subfloat[]{\includegraphics[width=1.6in]{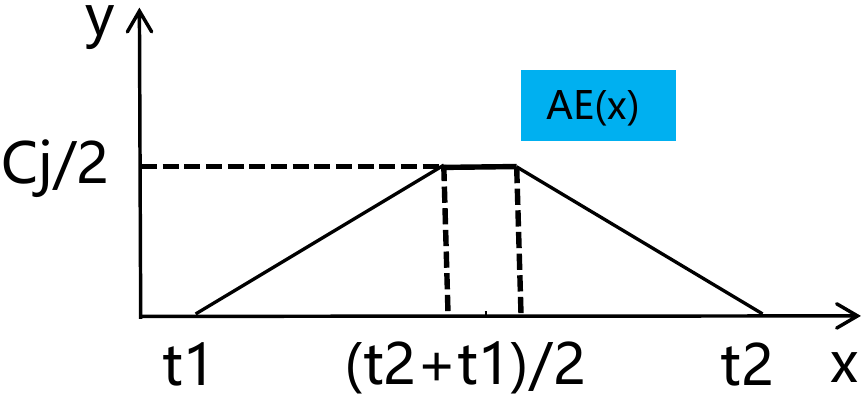}
		\label{fig4_1}}
	\hfil
	\subfloat[]{\includegraphics[width=1.6in]{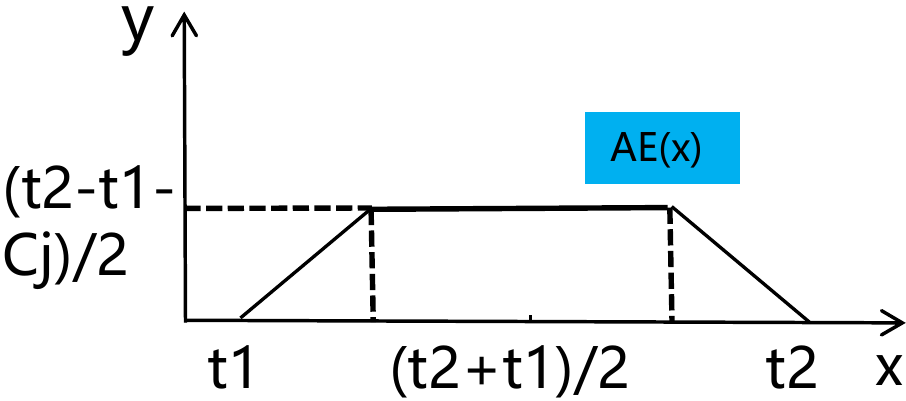}
		\label{fig4_2}}
	\caption{The upper bound of absolute error}
	\label{fig4}
\end{figure}

The query strategy of the FEH is especially effective for tracking elephant flows or large bursts of traffic. This is because when the number of 1's in the window is large, the time interval $\left[t_{1}+C_{j}, t_{2}-C_{j}\right]$ will be significantly shortened in the case of $C_{j} \leq \frac{1}{2}\left(t_{2}-t_{1}\right)$, and may further lead to the case of $C_{j}>\frac{1}{2}\left(t_{2}-t_{1}\right)$. Therefore, the accuracy will be improved more significantly when the query strategy of FEH used in elephant flows or large bursts of traffic. In addition, FEH accesses one more timestamp than EH when querying, thus the complexity of the query is still $O(1)$. The cost of adding the timestamp of the latest expired bucket is very small and will not affect the memory usage of the scheme. With the query speed and memory footprint almost unchanged, the query strategy of the EEH scheme further reduces the upper bound of absolute error to a minimum. Combining with the update strategy of the flattened count, the accuracy of FEH is further improved.

\subsection{The Practical design of FEH}
We design a practical structure of FEH, referred to as P-FEH. We regard the entire bucket array of the P-FEH model as a two-dimensional array with $\frac{k}{2}+1$ columns and $\log \left(\frac{2 N}{k}\right)+2$ rows, and each row is designed to be a circular array with a head and tail pointer. The P-FEH initially sets all buckets to zero, the head and tail pointer of each row point to the bucket of the first column. When a bucket is inserted, the tail pointer moves backwards and the head pointer always points to the oldest bucket of the row. When the head and tail pointers point to the same bucket and the size of the bucket is not zero, this means that the line is fully filled. Each row of the P-FEH corresponds to a different bucket size except that the bucket size=1 occupies two rows. The bucket insertion of each row is done by the tail pointer, and the bucket merging of each row is done by the head pointer.

Considering the efficiency of the FEH in practical application, we have slightly changed the update strategy of the P-FEH structure compared with FEH.
\subsubsection{The FEH merge operation} The FEH merge operation in P-FEH structure adopts a ``diligent'' merge method. We limit the column number $\frac{k}{2}+1$ to even number, and the merge operation of P-FEH is done in units of row. For one merge operation, the P-FEH combines the buckets of the oldest row of partition size, and half of that line will become empty buckets, as show in Fig.5(a) to Fig.5(b). We call this line ``P-FEH merge row''. After that, these empty buckets are not really placed next to the newest bucket of the partition size as FEH dose. We use another way to achieve the same effect, but it has better efficiency.

\subsubsection{The EH update mode after the FEH merge operation} Like FEH, after the FEH merge operation, the P-FEH will continue the EH update mode to complete the remaining update steps. However, there is a little different. As shown in Fig.5(b) to Fig.5(c), the P-FEH first merge the two oldest buckets in the oldest row of partition size, and write the merged buckets into the P-FEH merged line. Then it writes the bucket of partition size formed by the EH update process into the oldest row of partition size. At this point, the oldest row of the partition size has both the oldest and newest bucket of partition size, thus we call it ``new-old row''.

\begin{figure}[htbp]
	\centerline{\includegraphics[width=\linewidth]{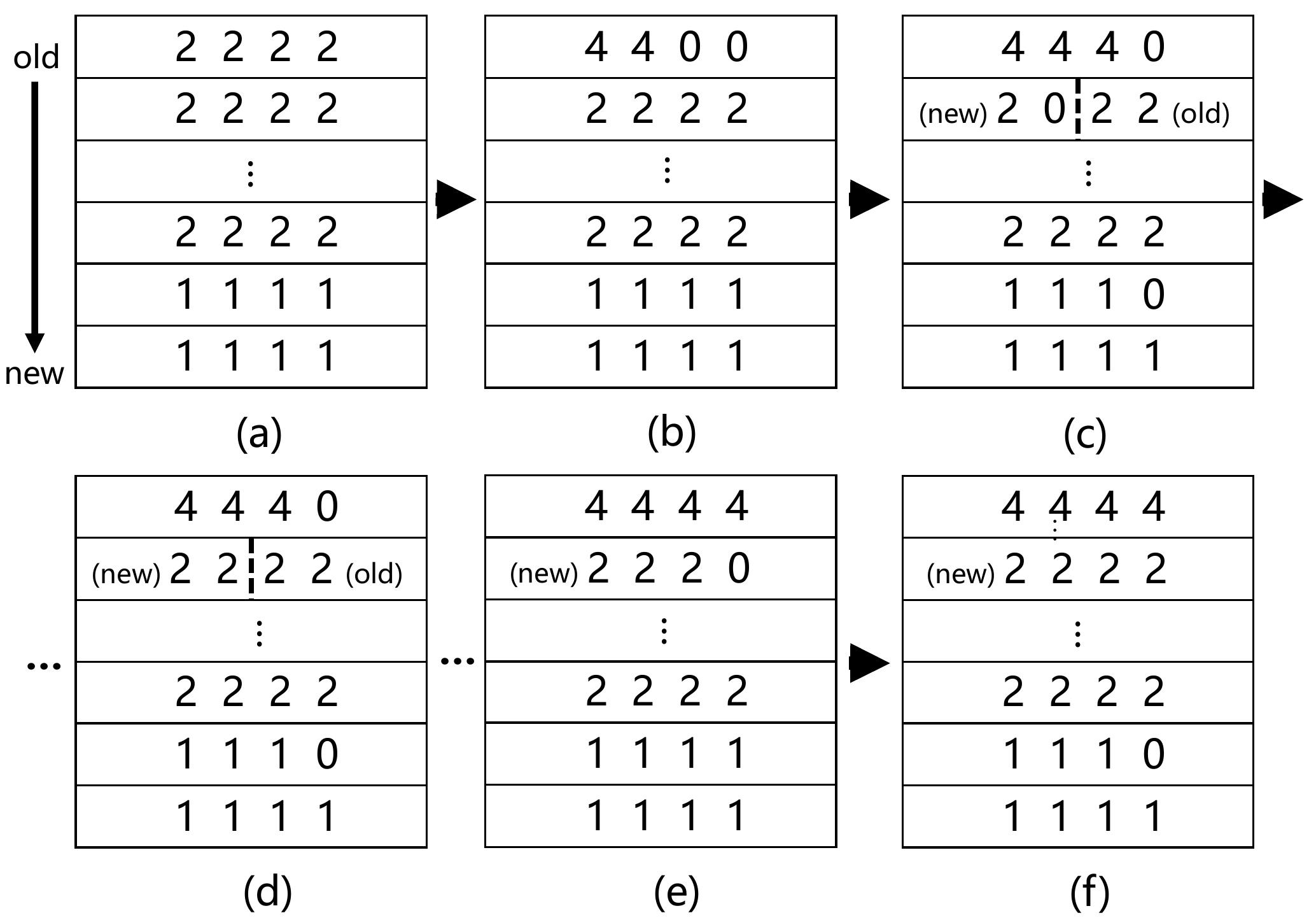}}
	\caption{The update of P-FEH}
	\label{fig5}
\end{figure}

\subsubsection{The EH update mode before the FEH merge operation} For the general EH update mode, after generate the bucket of partition size, if the bucket array has the new-old row of partition size, write the bucket of partition size formed by the EH update process to that row first(as shown in Fig.5(c) to Fig.5(e)), otherwise, write the bucket to the newest row of partition size(as shown in Fig.5(e) to Fig.5(f)), and if there are no empty bucket in this row, the P-FEH merge operation will be performed. When there are only new buckets in the new-old row, it becomes the newest row of the partition size. In order to maintain the orderliness between rows, we can move this newest row to the corresponding positions. However, this is less efficient, and we take another approach in P-FEH. 

In P-FEH, we use another structure to mark the order of each row, called tag structure. When updating, we look up this tag structure to update the corresponding rows. In applications, the size of the sliding window is often set at $2^{16} \sim 2^{30}$ [26], [34], [36]. Therefore, we can use the word acceleration technology to limit the tag structure to one or several machine words. It decreases the number of memory accesses, further reduce the cost of update.

\subsubsection{The processing of expired buckets} If there is an expired bucket in the oldest row in the P-FEH structure and the buckets of this row are not all expired, in order to ensure that the P-FEH merge operation is performed in units of row, this row will not participate in P-FEH update. This row will become the newest row of its original bucket size after its buckets have all expired, and will be used for future EH update mode. Moreover, because there is the new-old row in the P-FEH structure, it is important to note that when the older buckets in the new-old row expire, the new-old row will become the newest row of its size.

Other parts of the update strategy and the query strategy of the P-FEH are the same as those in FEH, and the final update effect is also the same as FEH. 

\section{EXPERIMENTAL RESULTS}
In this section, we first evaluate the performance of the P-FEH and the EH on two real datasets. Then, in order to show the relationship between the improvement of accuracy of the FEH query strategy and the number of 1's in the window, we generate a set of synthetic datasets, in which each dataset has a different proportion of 1's in the window.
\subsection{Metrics}
\textbf{Average Absolute Error \ (AAE):} AAE is defined as $\frac{1}{|N|} \sum_{i=1}^{N}\left|f_{i}-\hat{f}_{i}\right|$ , where $f_{i}$ is the real frequency of the $i$th element, $\hat{f}_{i}$ is the estimated frequency of this element, and $N$ is the total number of distinct elements in the query set. 

\textbf{Average Relative Error \ (ARE):}  ARE is defined as  $\frac{1}{|N|} \sum_{i=1}^{N}\left|f_{i}-\hat{f}_{i}\right| / f_{i}$, where $f_i$ is the real frequency of the $i$th element, $\hat{f}_{i}$ is the estimated frequency of this element, and $N$ is the total number of distinct elements in the query set. 

\textbf{Average Absolute Error Average Upper Bound \ (AAEUB):} Average absolute upper bound is calculated by $\frac{C_{j}}{2}$, where $\mathcal{C}_{j}$ is the oldest bucket size. The AAEUB is the average upper bound of the absolute error for each query.

\textbf{Average Relative Error Upper Bound \ (AREUB):}Average error upper bound is calculated by \eqref{eq2}. The AREUB is the average upper bound of the relative error for each query.

\textbf{Throughput:} We simulate the actual update and query process of the EH and P-FEH model on the CPU platform and calculate the throughput using mega-instructions per second (Mips). 

\subsection{Experimental Setup}
\subsubsection{Dataset}
\paragraph{CAIDA Dataset}This dataset is from CAIDA Anonymized Internet Trace 2016 [43], which contains IP packets. Consider packages with the same source and destination IP addresses as the same item, the trace has 10M packets and around 4.2M distinct items. 
\paragraph{Campus Dataset} This dataset is from the real DNS trace of our campus from 10:00 am~11:00 am on October 19, 2018. Consider packages that request and respond to the same domain name as the same item, the trace has 10M packets and around 5.3M items.
\paragraph{Synthetic Datasets} The synthetic datasets consist of 8 datasets containing only 0's and 1's. The same is that each dataset contains 1M elements, while the difference is that the proportion of 1's in each dataset is different, ranging from 0.02 to 0.72 with a step of 0.1.

\begin{figure*}[!t]
	\centering
	\subfloat[AAE on CAIDA dataset vs. k]{\includegraphics[width=1.75in]{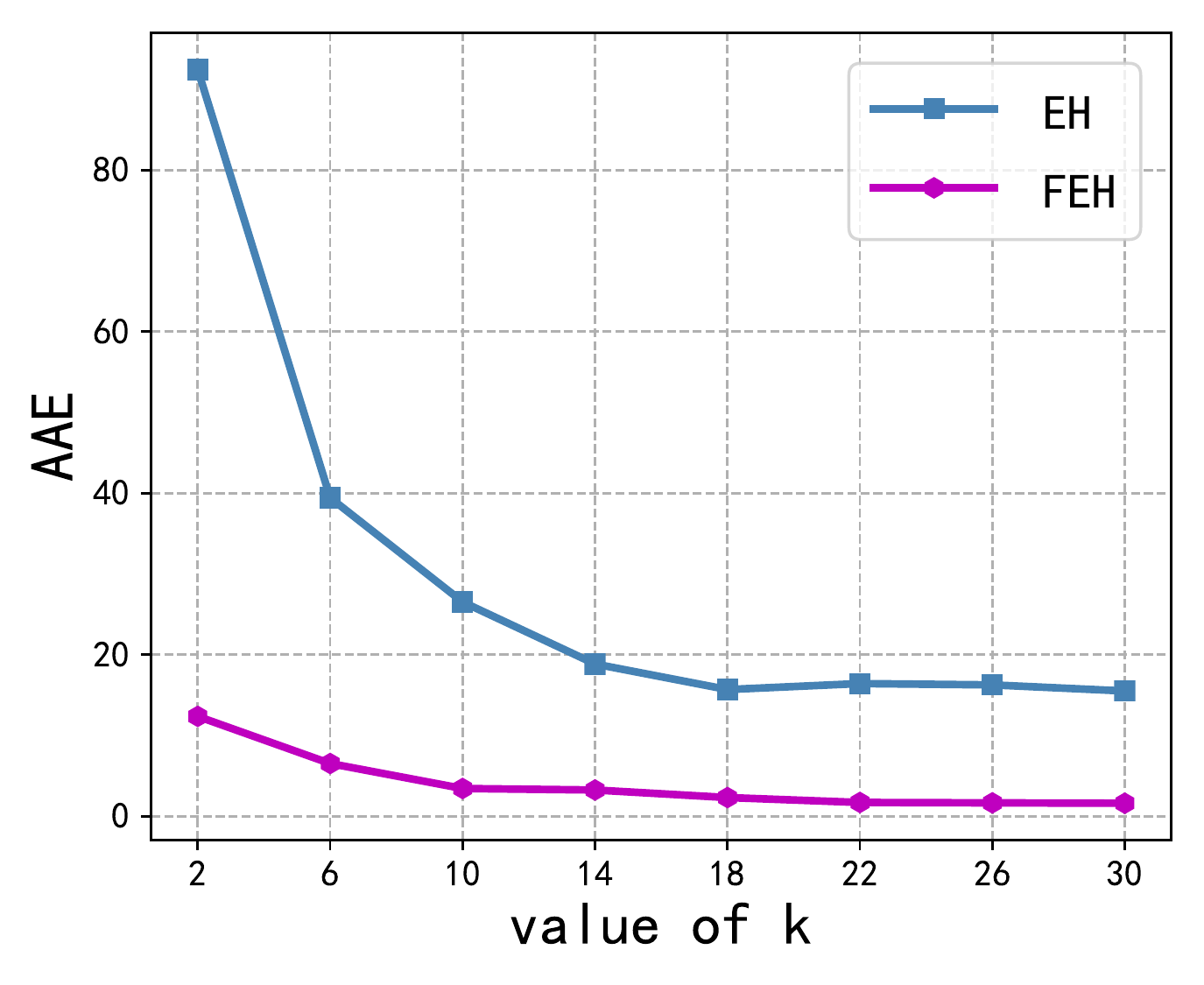}
		\label{fig6_3}}
	%	\hfil
	\subfloat[ARE on CAIDA dataset vs. k]{\includegraphics[width=1.75in]{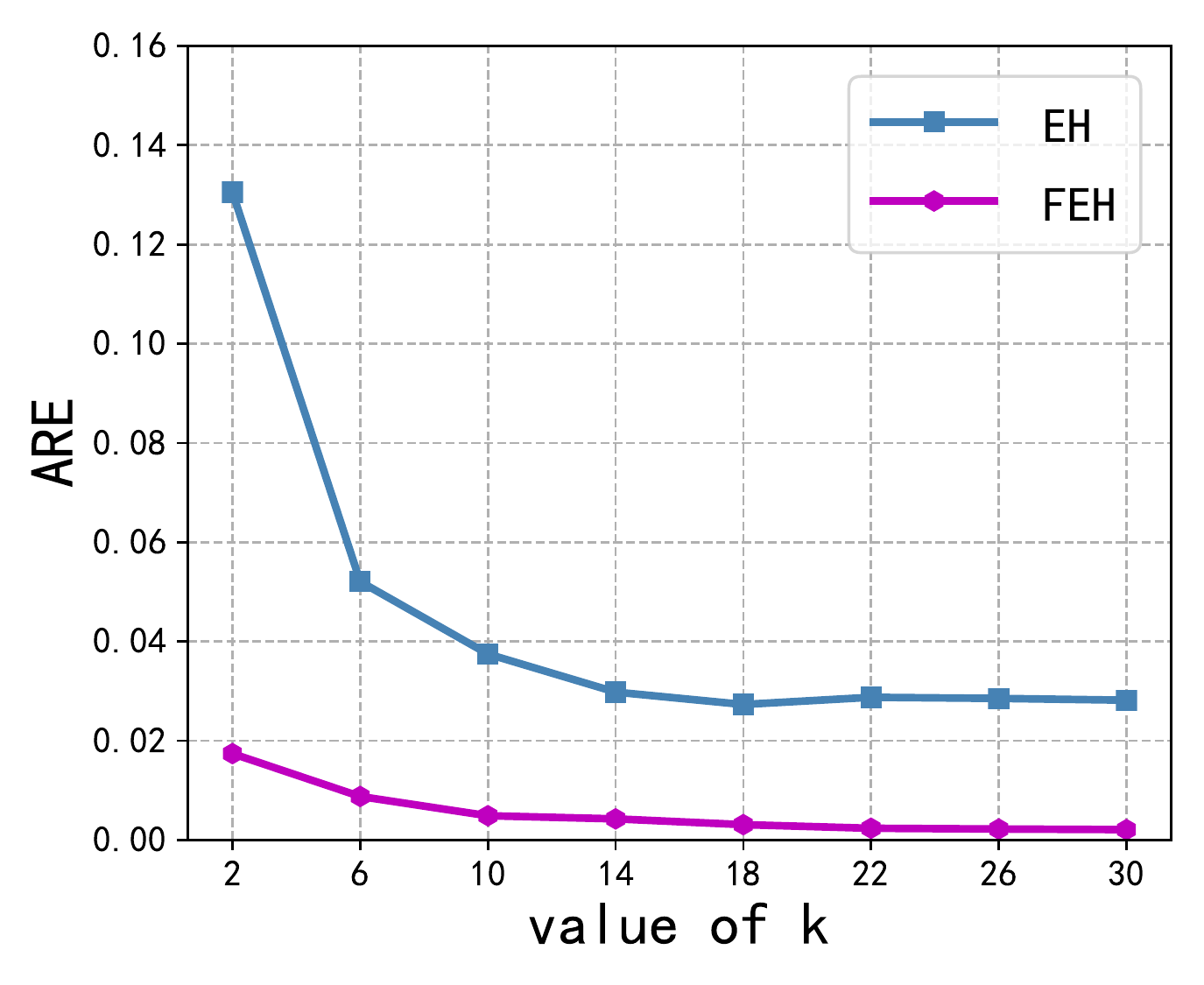}
		\label{fig6_4}}
	%	\hfil
	\subfloat[AAE on campus dataset vs. k]{\includegraphics[width=1.75in]{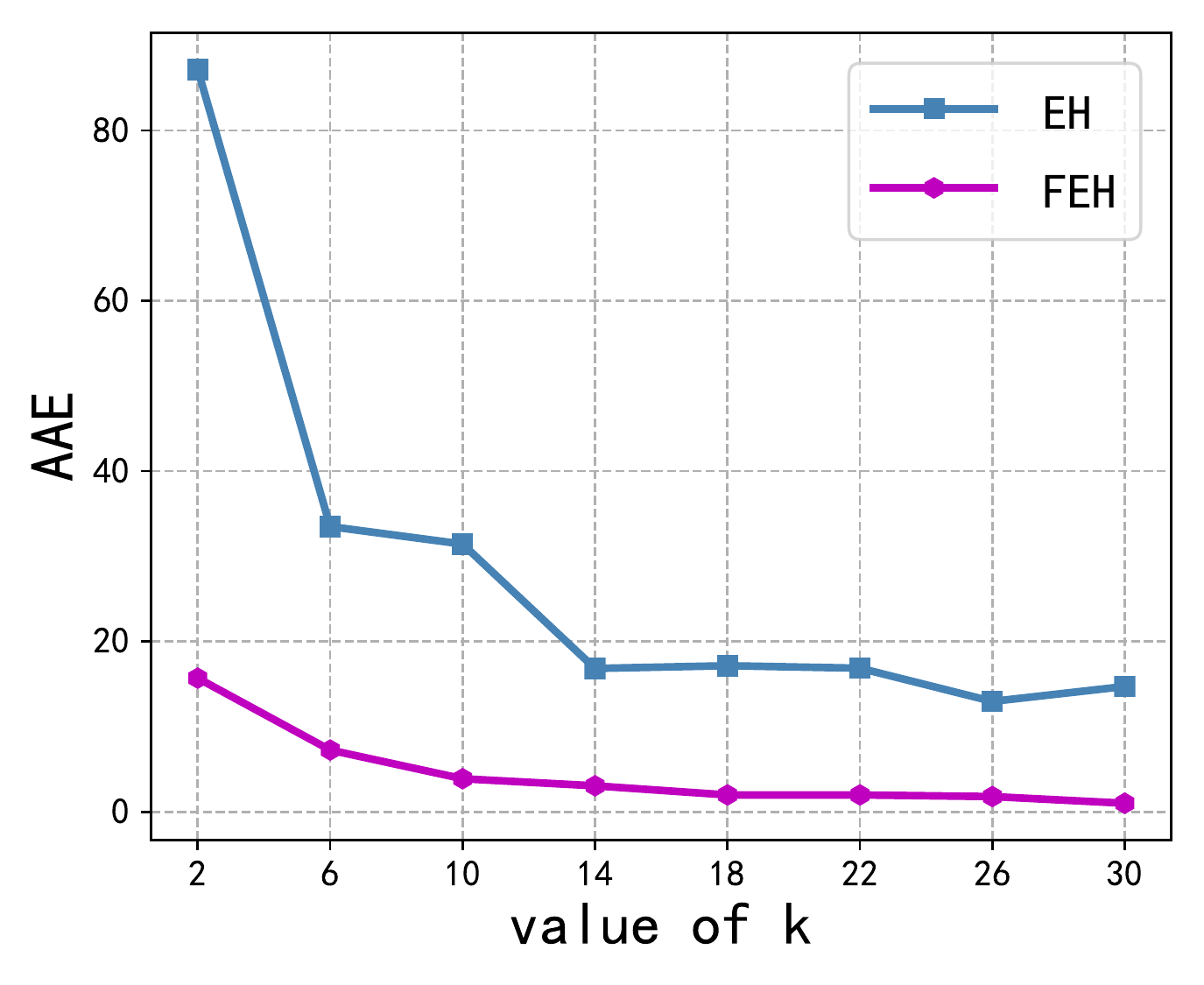}
	\label{fig6_1}}
	%	\hfil
	\subfloat[ARE on campus dataset vs. k]{\includegraphics[width=1.75in]{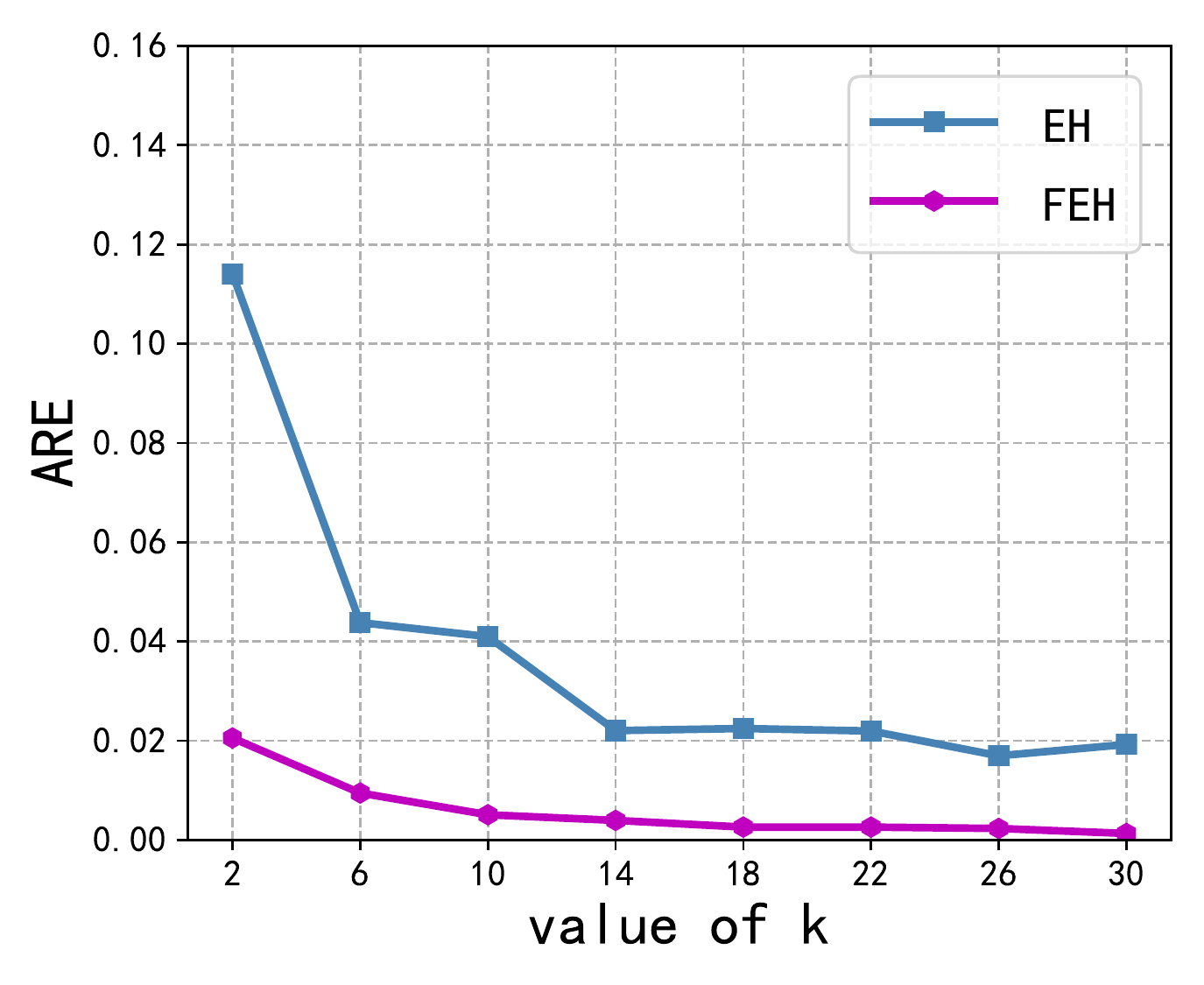}
	\label{fig6_2}}
	%	\hfil
	\caption{AAE and ARE of EH and P-FEH with different k on real world datasets}
	\label{fig6}
\end{figure*}

\begin{figure*}[!t]
	\centering
	\subfloat[AAE on CAIDA dataset vs. window size]{\includegraphics[width=1.75in]{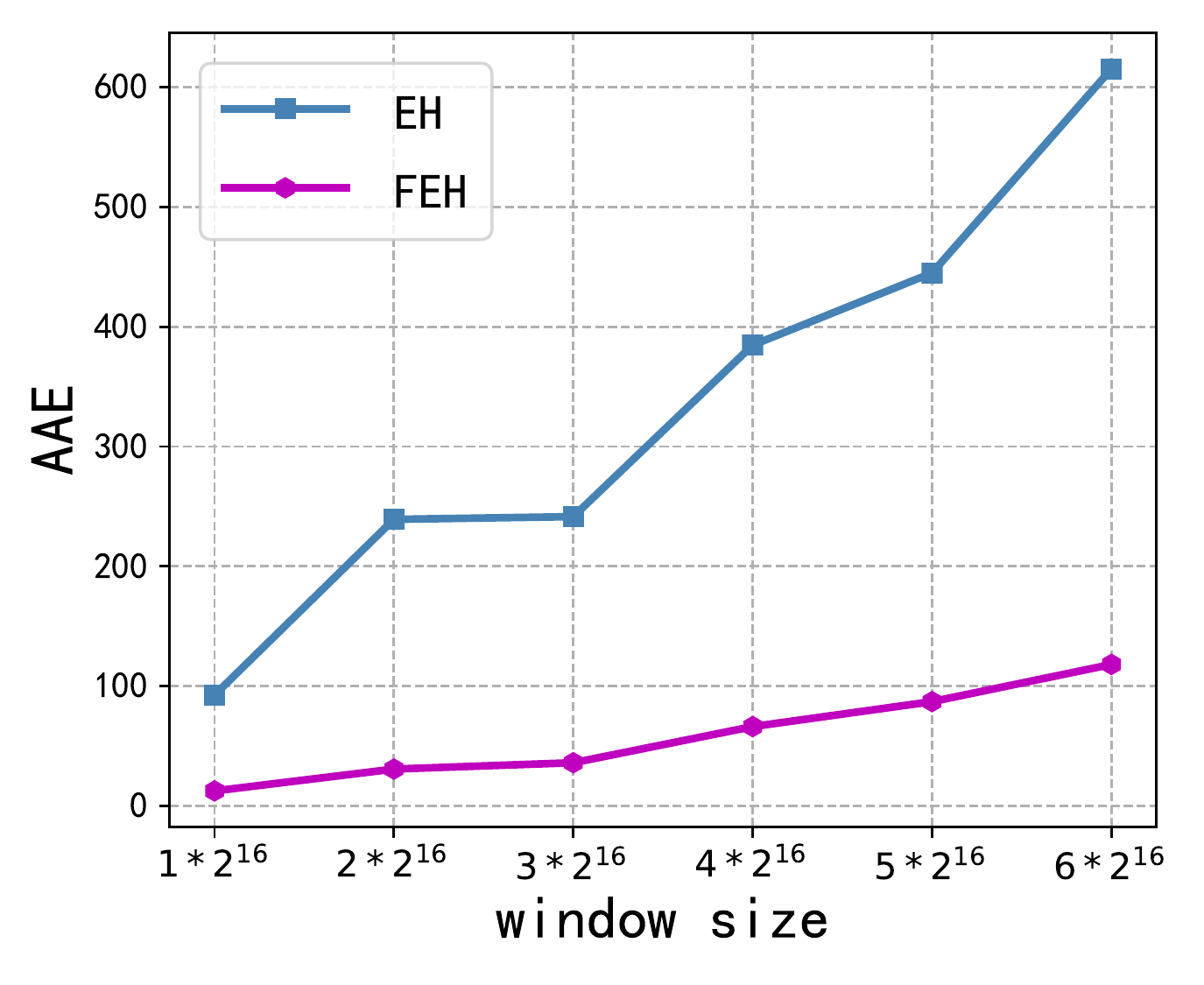}
		\label{fig7_3}}
	%	\hfil
	\subfloat[ARE on CAIDA dataset vs. window size]{\includegraphics[width=1.75in]{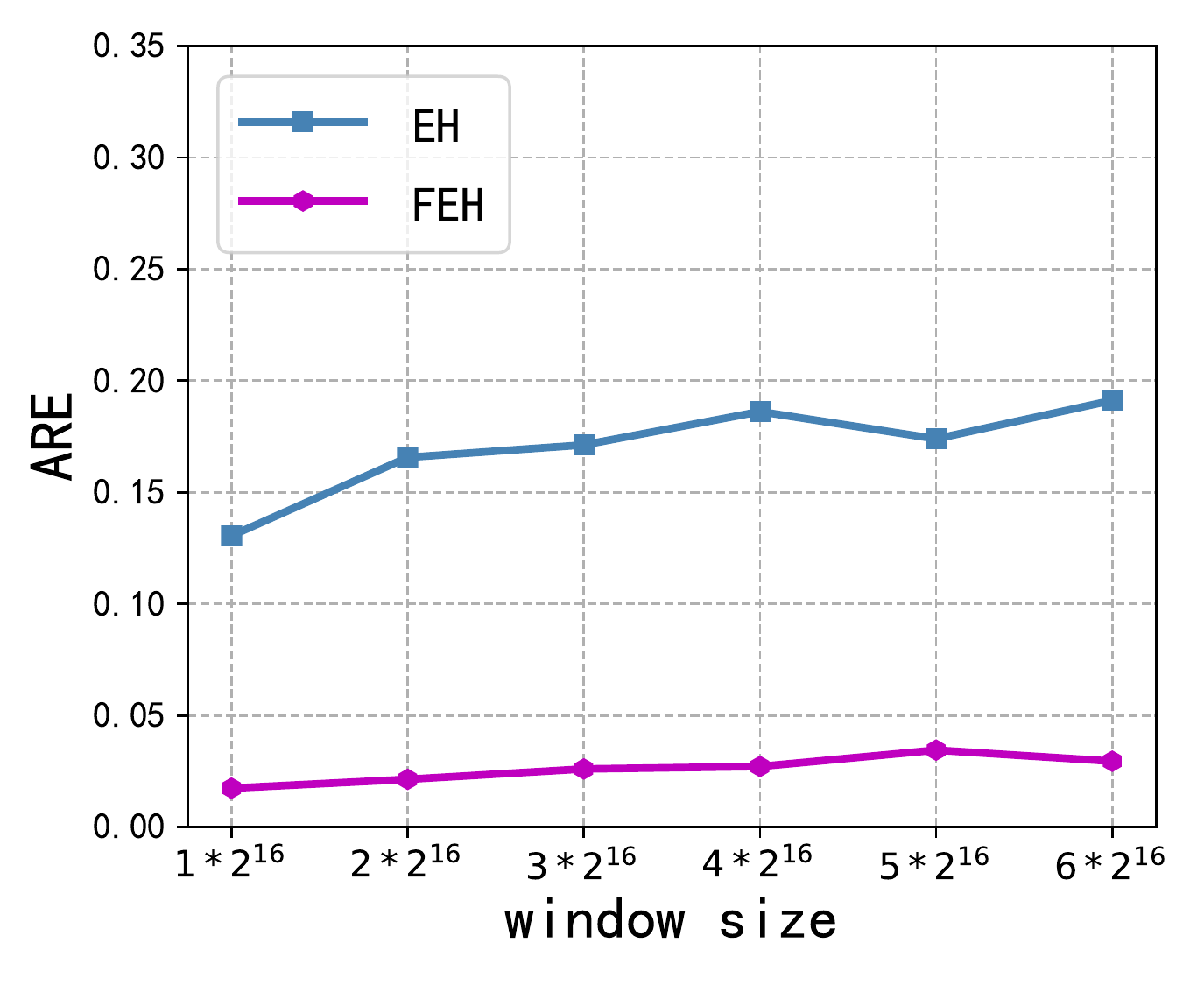}
		\label{fig7_4}}
	%	\hfil
	\subfloat[AAE on campus dataset vs. window size]{\includegraphics[width=1.75in]{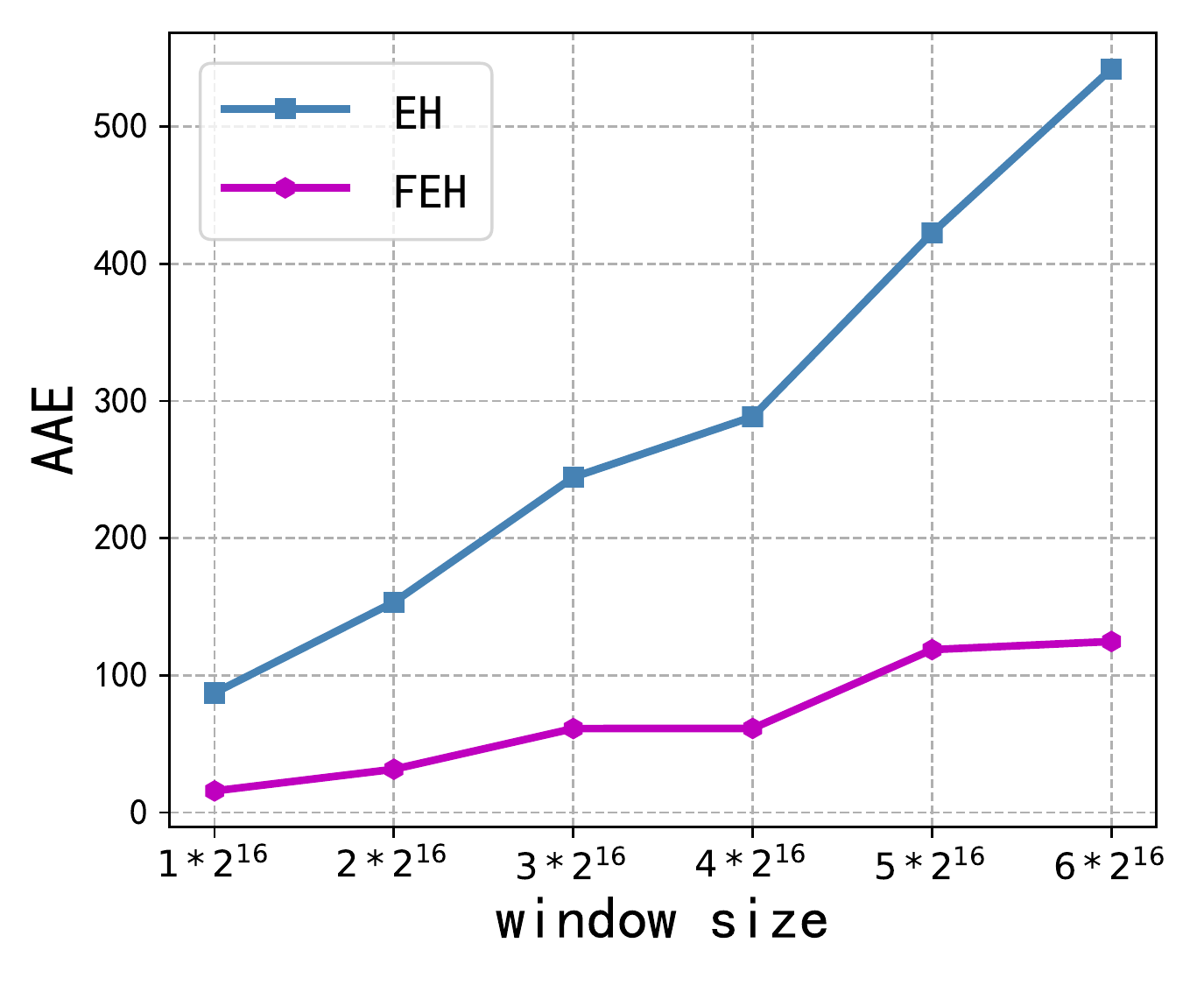}	
	\label{fig7_1}}
	%	\hfil
	\subfloat[ARE on campus dataset vs. window size]{\includegraphics[width=1.75in]{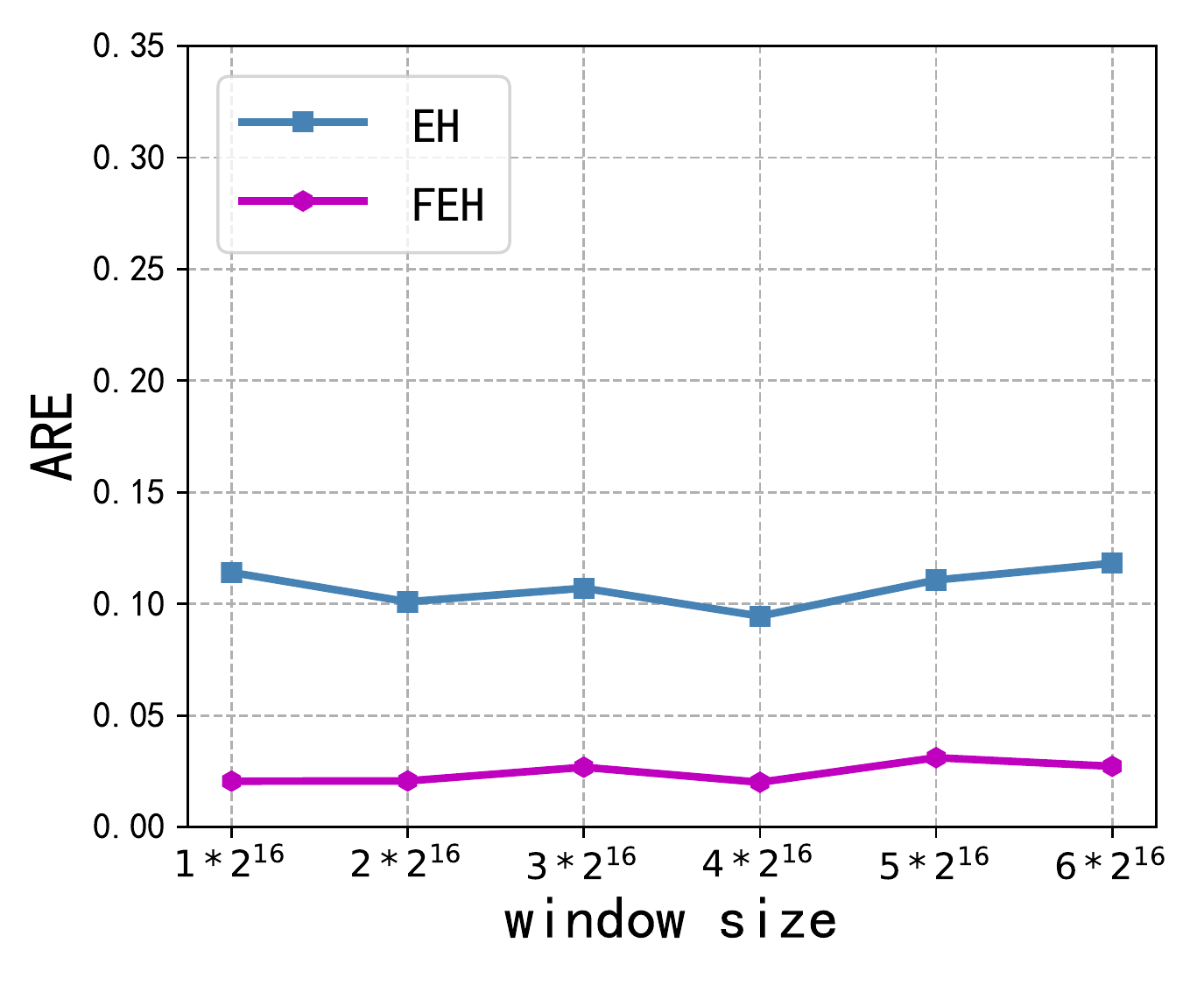}
	\label{fig7_2}}
	%	\hfil
	\caption{AAE and ARE of EH and P-FEH with different window size on real world datasets}
	\label{fig7}
\end{figure*}

\begin{figure*}[!t]
	\centering
	\subfloat[Update throughput on CAIDA dataset vs. k]{\includegraphics[width=1.75in]{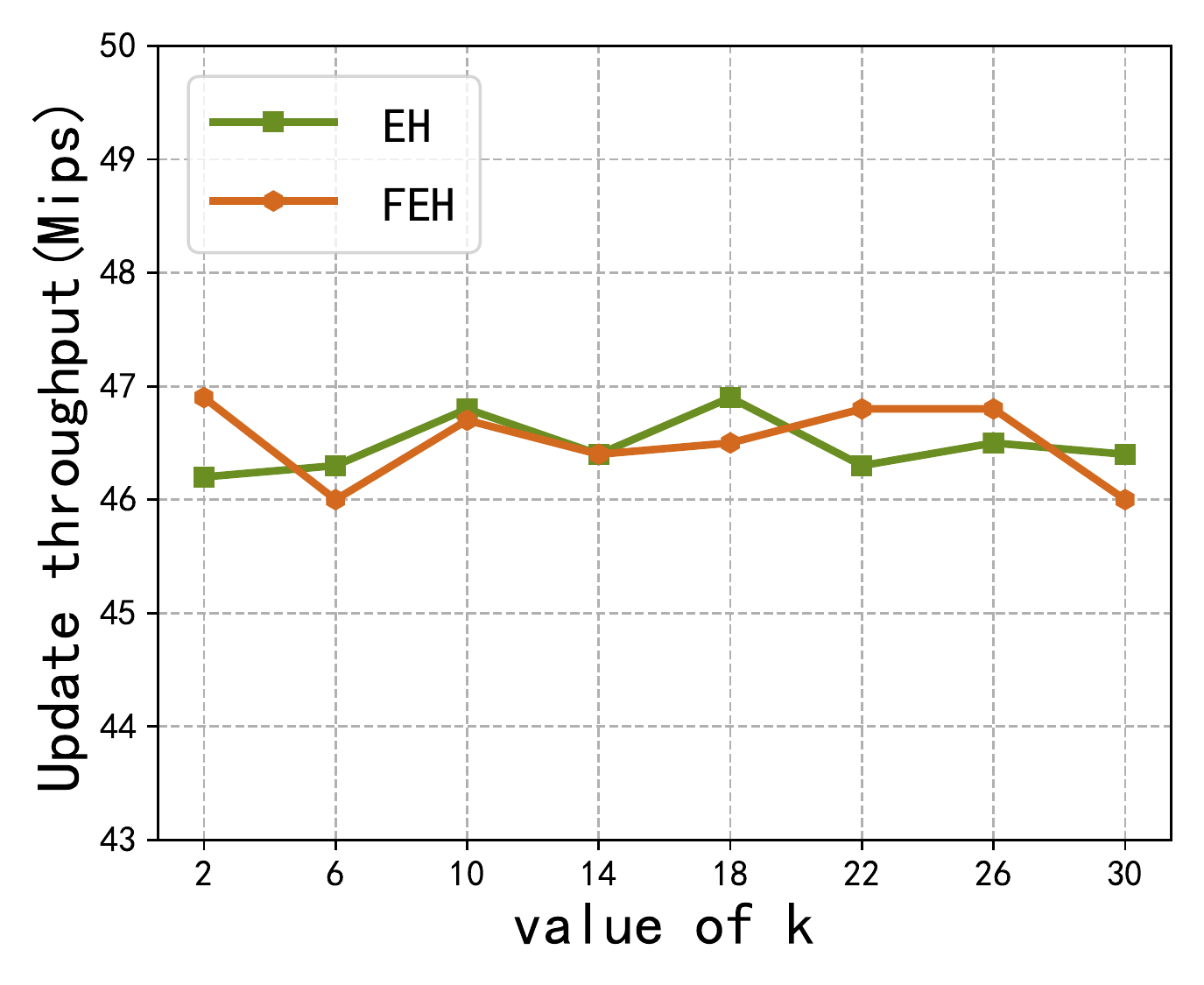}
		\label{fig8_1}}
	%	\hfil
	\subfloat[Query throughput on CAIDA dataset vs. k]{\includegraphics[width=1.75in]{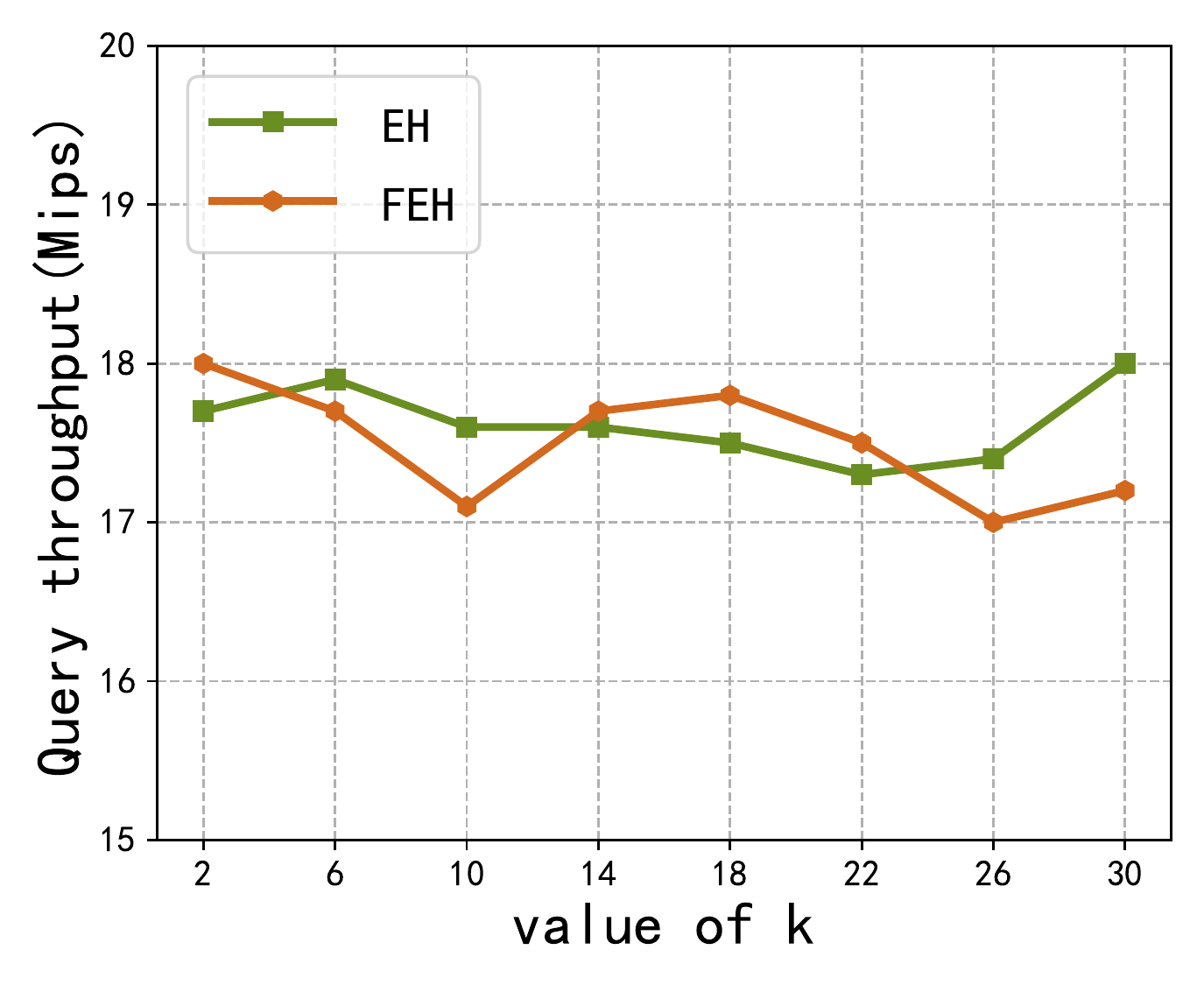}
		\label{fig8_2}}
	\subfloat[Update throughput on campus dataset vs. k]{\includegraphics[width=1.75in]{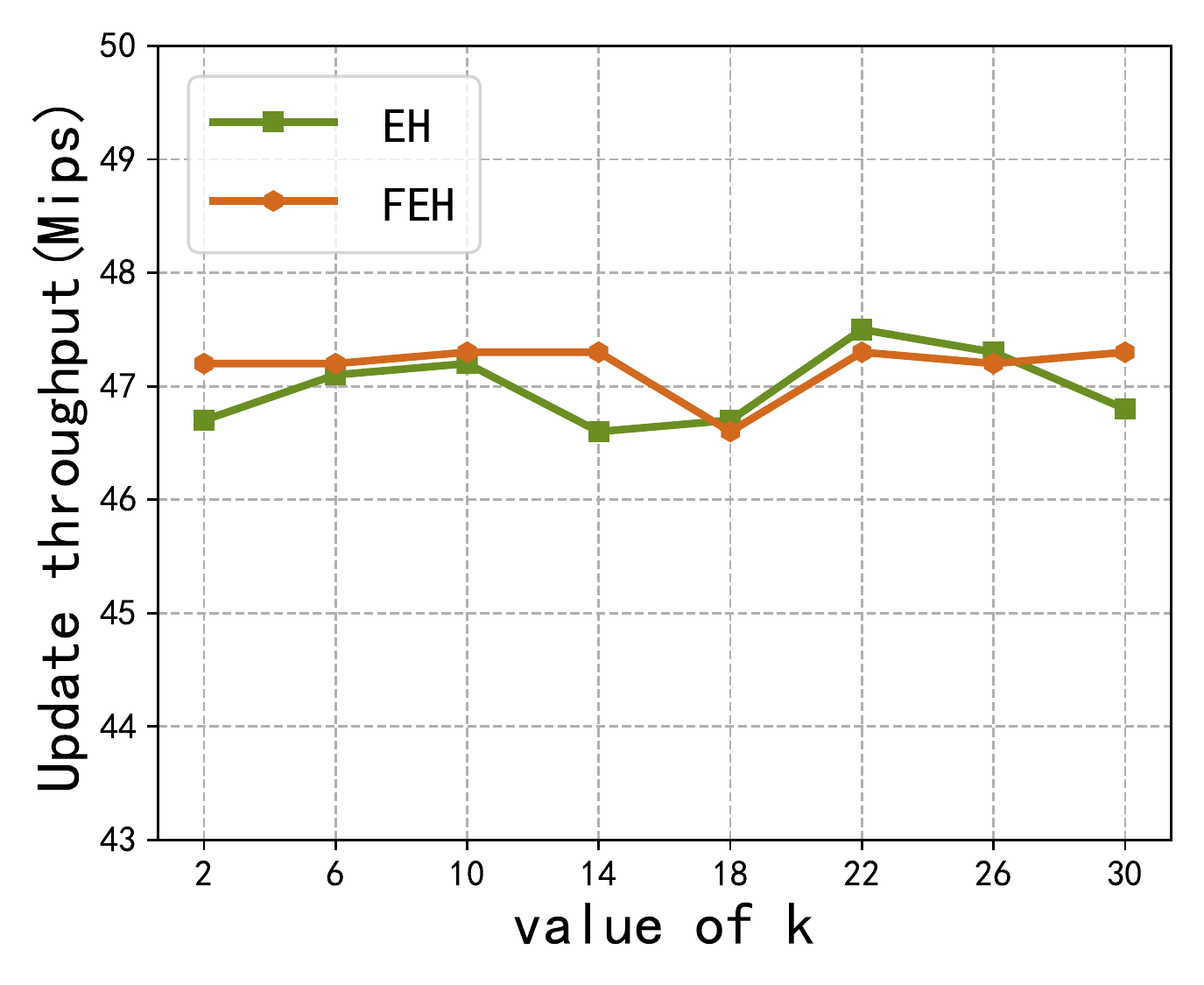}
	\label{fig8_3}}
	%	\hfil
	\subfloat[Query throughput on campus dataset vs. k]{\includegraphics[width=1.75in]{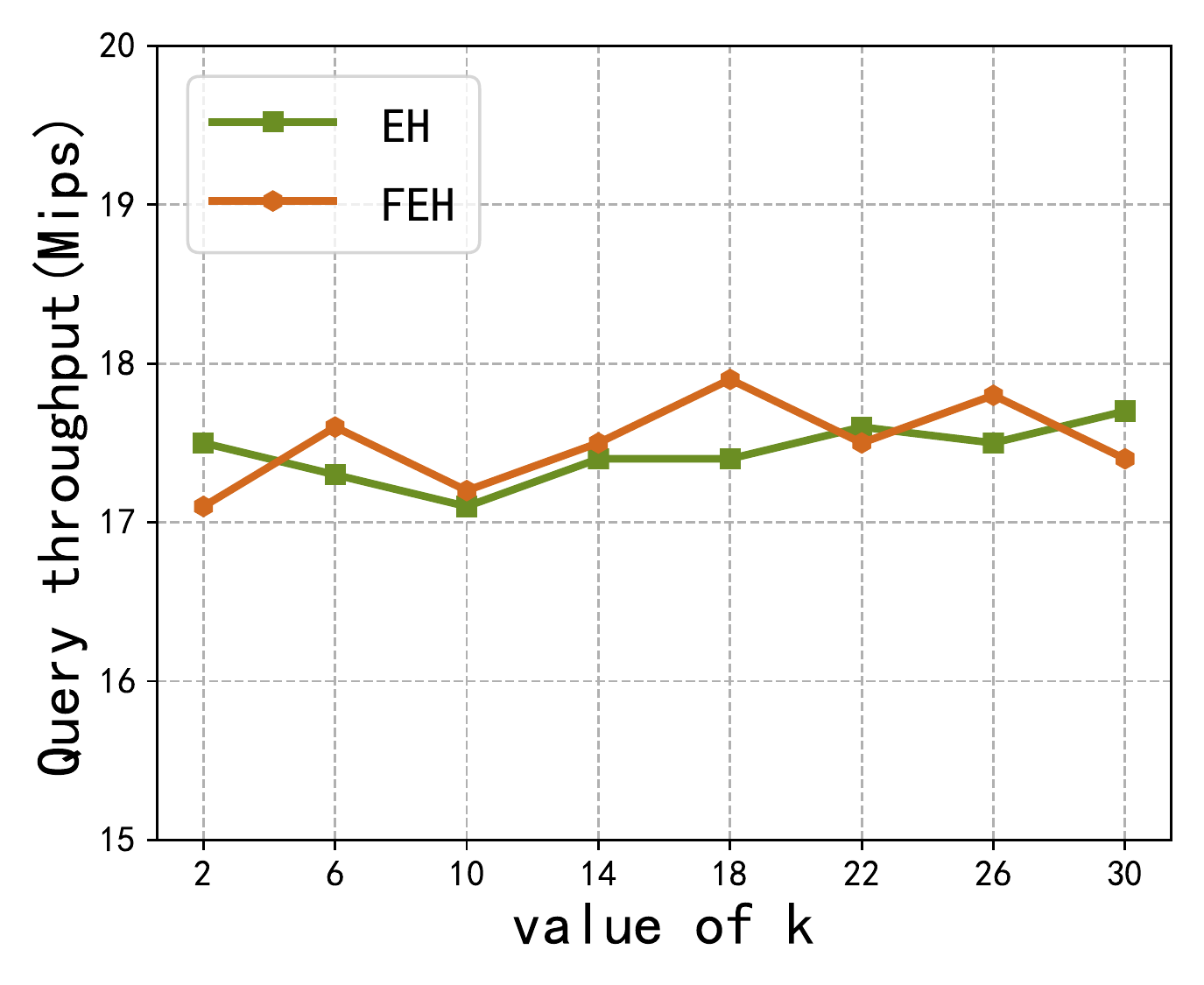}
	\label{fig8_4}}
	%	\hfil
	\caption{Update throughput and Query throughput of EH and P-FEH with different k on real world datasets}
	\label{fig8}
\end{figure*}

\subsubsection{Implementation}
We have implemented the EH and the P-FEH in C++. Like P-FEH, the bucket array of EH is also a two-dimensional array with a head and tail pointer per row. The memory allocation of the bucket array for both models is based on (3), and we let $N^{\prime}=window \ size$. In PEH, we use three additional machine words to accommodate the tag structure and the extra timestamp used in the query strategy. Hence, the memory occupancy of the EH and the P-FEH can be regarded as the same. In all our experiments, we let the sliding window sliding from the beginning to the end of the datasets, and the update throughput is exactly the average throughput of all updates in this process. Besides, we randomly selected 10000 time points in this process to perform query operations. For the CAIDA and the campus dataset, the Basic Counting problem corresponds to estimating the frequency of one specified item in the sliding window, which is equivalent to treating this item as `1' and other items as `0'. In the experiments on these two datasets, we randomly select 100 distinct items in multiple frequency bands as the measurement targets, and the error and the query throughput results are the average of these 100*10000 experimental results. For the synthetic datasets, they only contain 0's and 1's, thus the error and the query throughput results are the average of the 10000 experimental results. In all our experiments, unless noted otherwise, the window size is $2^{16}$ and the parameter $k$ is $2$ by default. We performed all the experiments on a machine with 2-core CPUs (2 threads, Pentium(R) Dual-Core CPU E5800 @3.2 GHz) and 4 GB total DRAM memory.

\subsection{The Accuracy of EH and P-FEH}
Fig.6 plots the AAE and the ARE of the EH and the P-FEH on different $k$ increasing from 2 to 30 with a step of 4. For the CAIDA dataset, the AAE of P-FEH is between 5.8 to 9.8 and on average 7.8 times smaller than that of the EH, and the ARE of P-FEH is between 5.9 to 13.5 and on average 8 times smaller than that of the EH. For the campus dataset, the AAE of P-FEH is between 4.6 to 14.7 and on average 7.9 times smaller than that of the EH, and the ARE of P-FEH is between 4.6 to 14.8 and on average 8 times smaller than that of the EH. The experimental results show that with the increase of $k$, the statistical accuracy of these two models improves, and the accuracy of P-FEH is always significantly higher than that of the EH. Fig.7 plots the AAEs and the AREs of the EH and the P-FEH on different window size increasing from $2^{16}$ to $6^{*} 2^{16}$ with a step of $2^{16}$. For the CAIDA dataset, the AAE of P-FEH is between 5.1 to 7.8 and on average 6.7 times smaller than that of the EH, and the ARE of P-FEH is between 5.1 to 7.8 and on average 6.4 times smaller than that of the EH. For the campus dataset, the AAE of P-FEH is between 4.1 to 5.6 and on average 5.2 times smaller than that of the EH, and the ARE of P-FEH is between 4.1 to 5.6 and on average 5.1 times smaller than that of the EH. The experimental results show that with the increase of window size, the accuracy of P-FEH is always significantly higher than that of the EH. Moreover, the AAE and the total count of these two models increase simultaneously, thus the ARE of these two models remains relatively stable. Our experimental results of accuracy have well proved that when facing different statistical requirements, the flattened count update strategy and the optimal query strategy of P-FEH can greatly improve the accuracy of sliding window queries.

\begin{figure*}[!t]
	\centering
	\subfloat[Update throughput on CAIDA dataset vs. window size]{\includegraphics[width=1.75in]{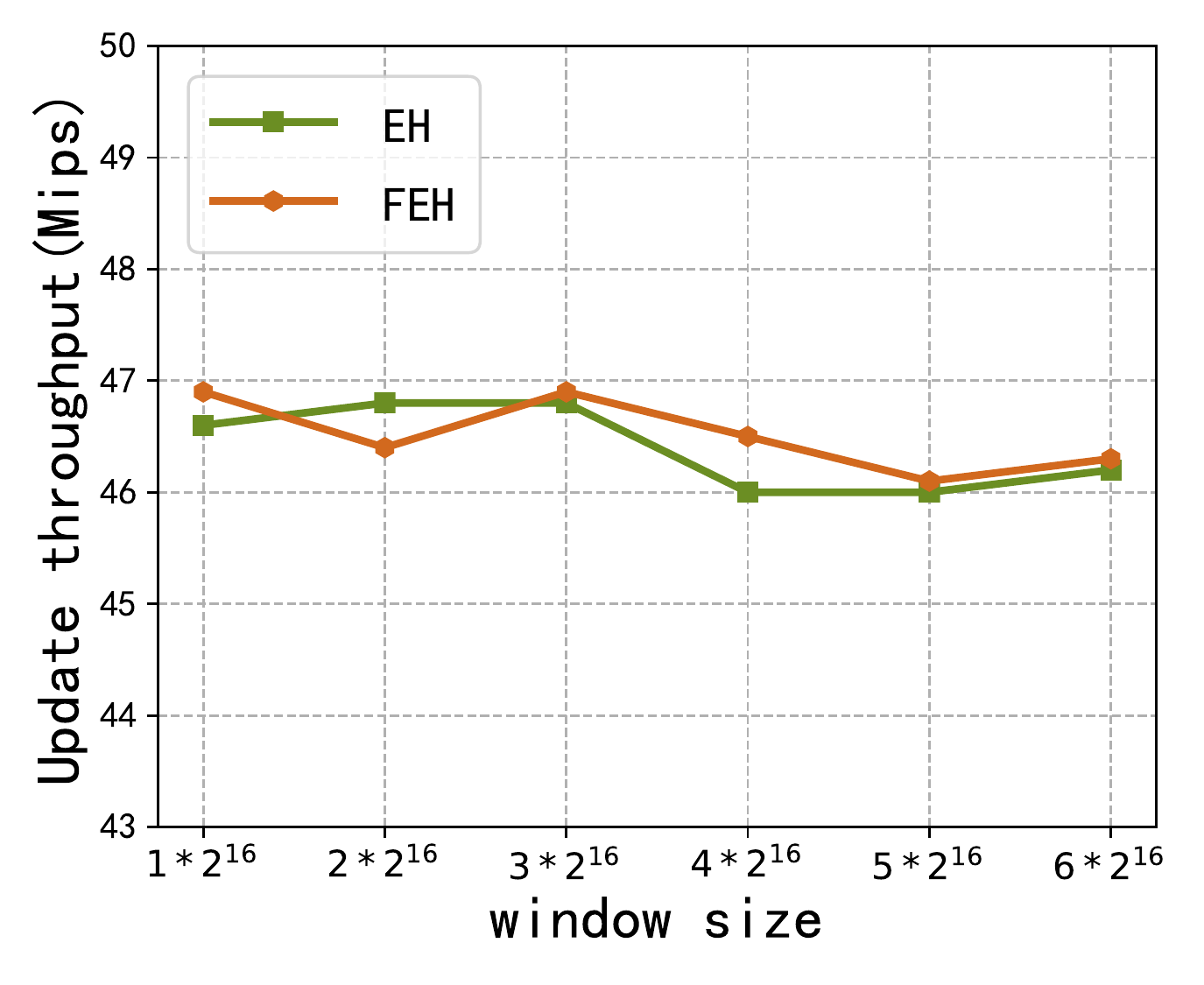}
		\label{fig9_5}}
	%	\hfil
	\subfloat[Query throughput on CAIDA dataset vs. window size]{\includegraphics[width=1.75in]{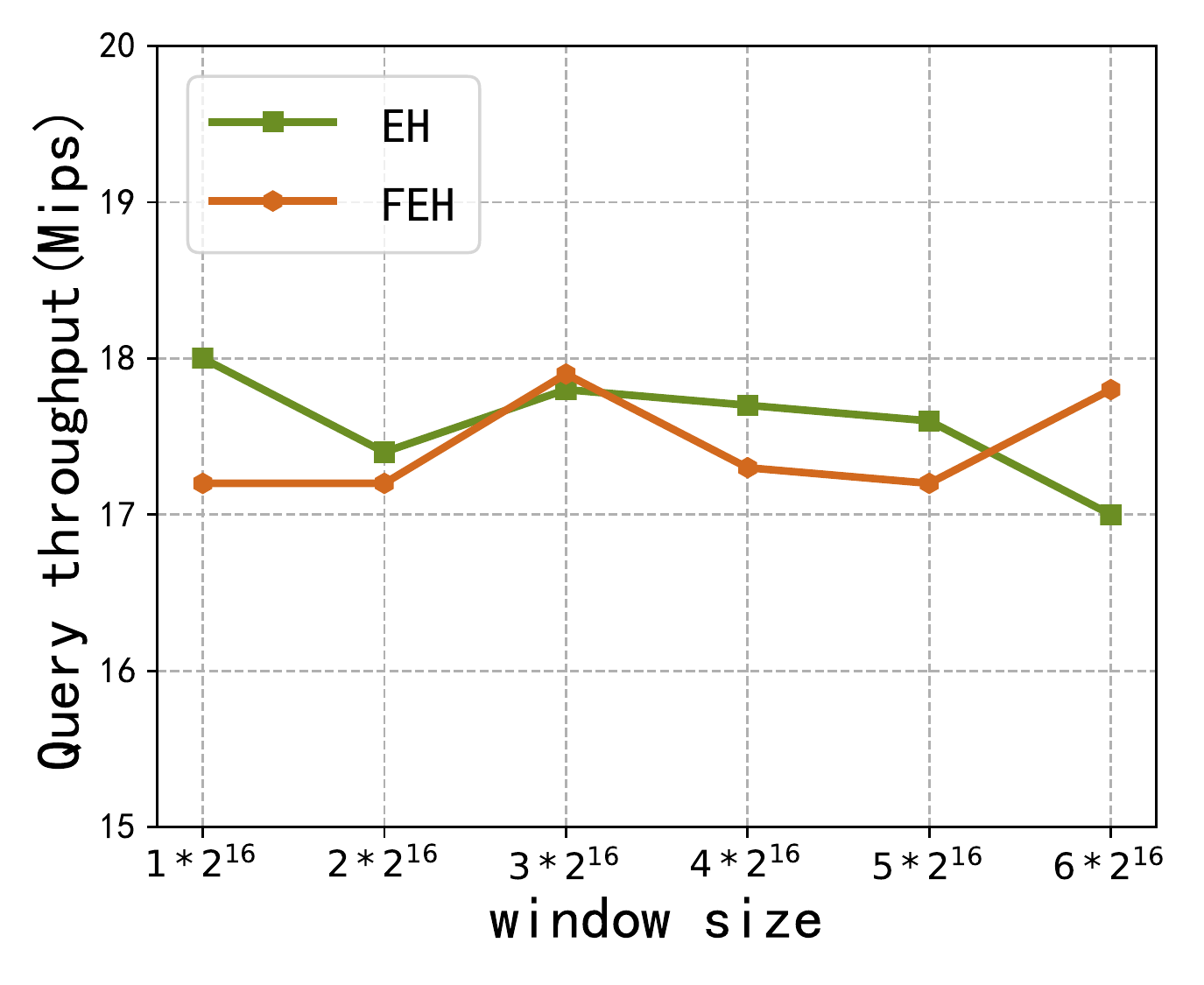}
		\label{fig9_6}}
	%	\hfil
	\subfloat[Update throughput on campus dataset vs. window size]{\includegraphics[width=1.75in]{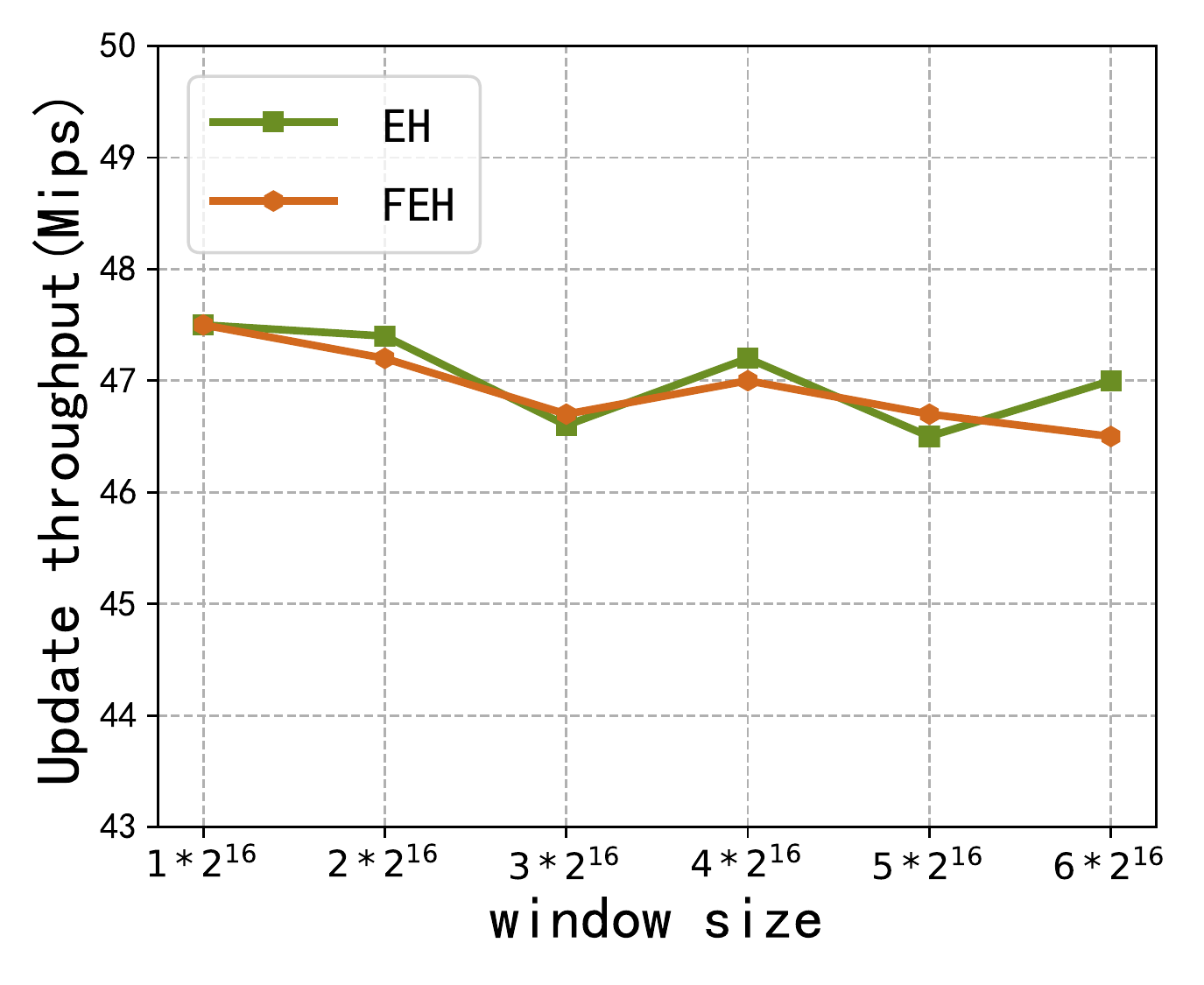}
	\label{fig9_7}}
	%	\hfil
	\subfloat[Query throughput on campus d
	ataset vs. window size]{\includegraphics[width=1.75in]{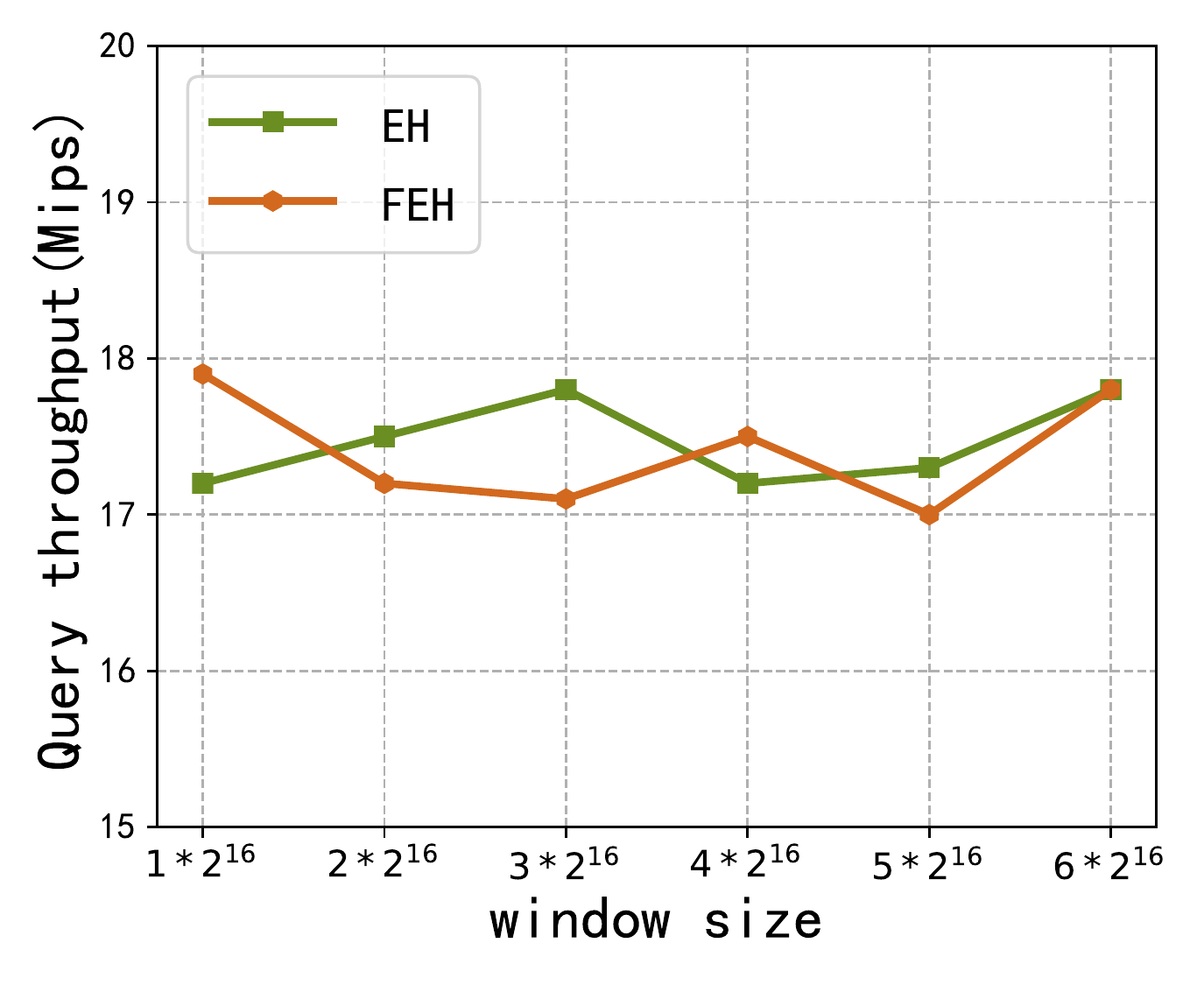}
	\label{fig9_8}}
	%	\hfil
	\caption{Update throughput and Query throughput of EH and P-FEH with different k on real world datasets}
	\label{fig9}
\end{figure*}

\begin{figure*}[!t]
	\centering
	\subfloat[AAE on campus dataset vs. percentage of 1 in the window]{\includegraphics[width=1.75in]{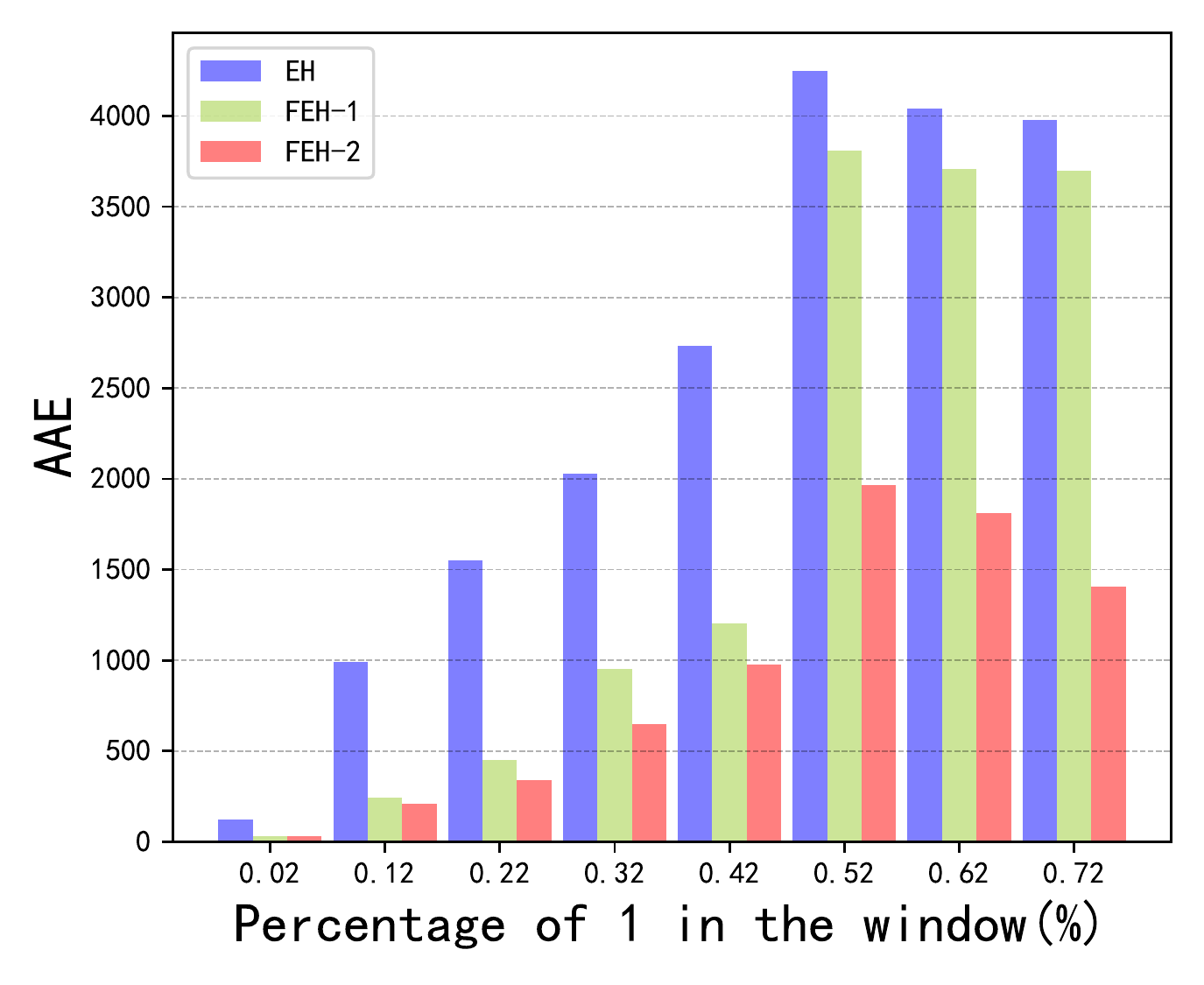}
		\label{fig10_1}}
	%	\hfil
	\subfloat[ARE on campus dataset vs. percentage of 1 in the window]{\includegraphics[width=1.75in]{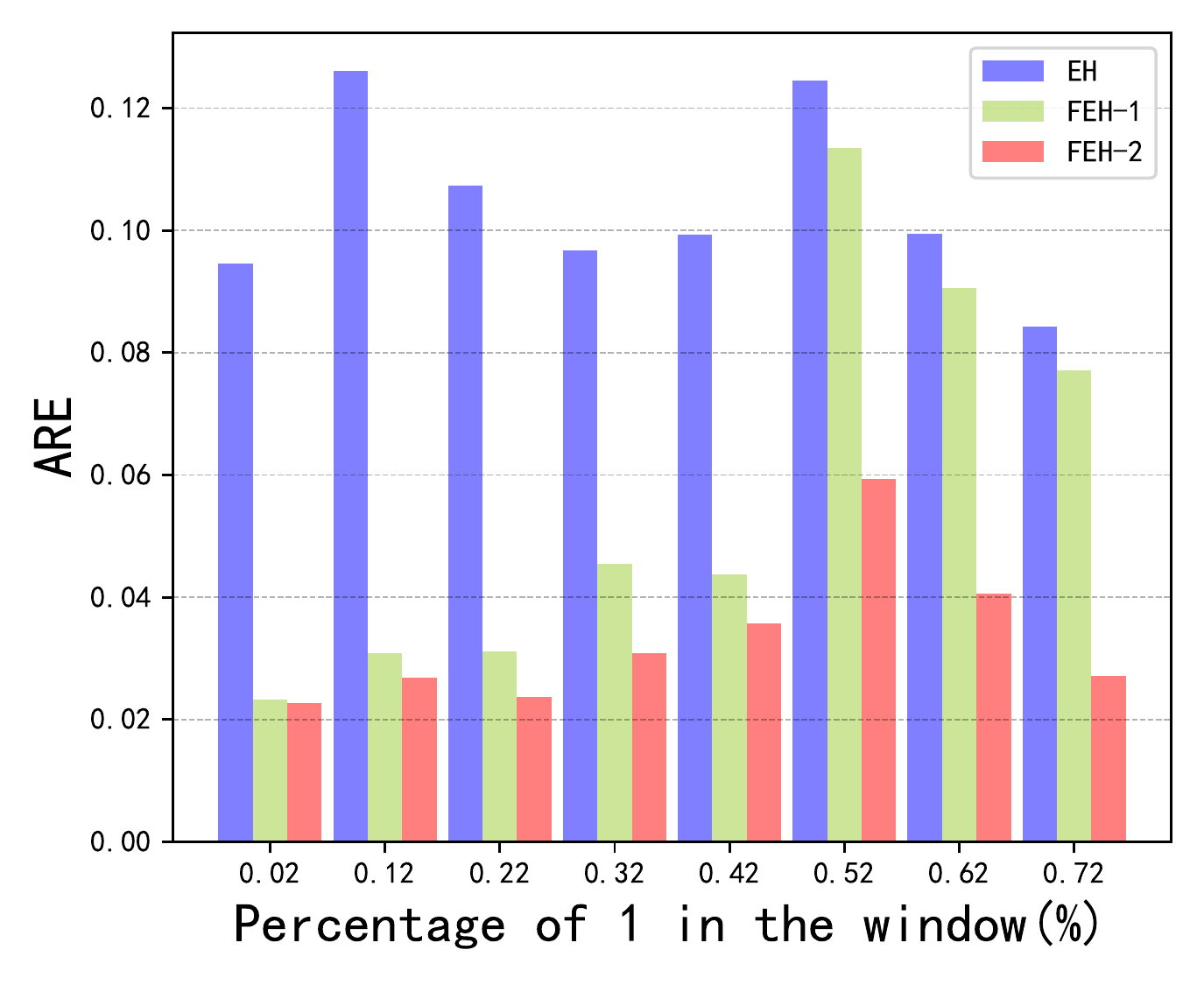}
		\label{fig10_2}}
	%	\hfil
	\subfloat[AAEUB on campus dataset vs. percentage of 1 in the window]{\includegraphics[width=1.75in]{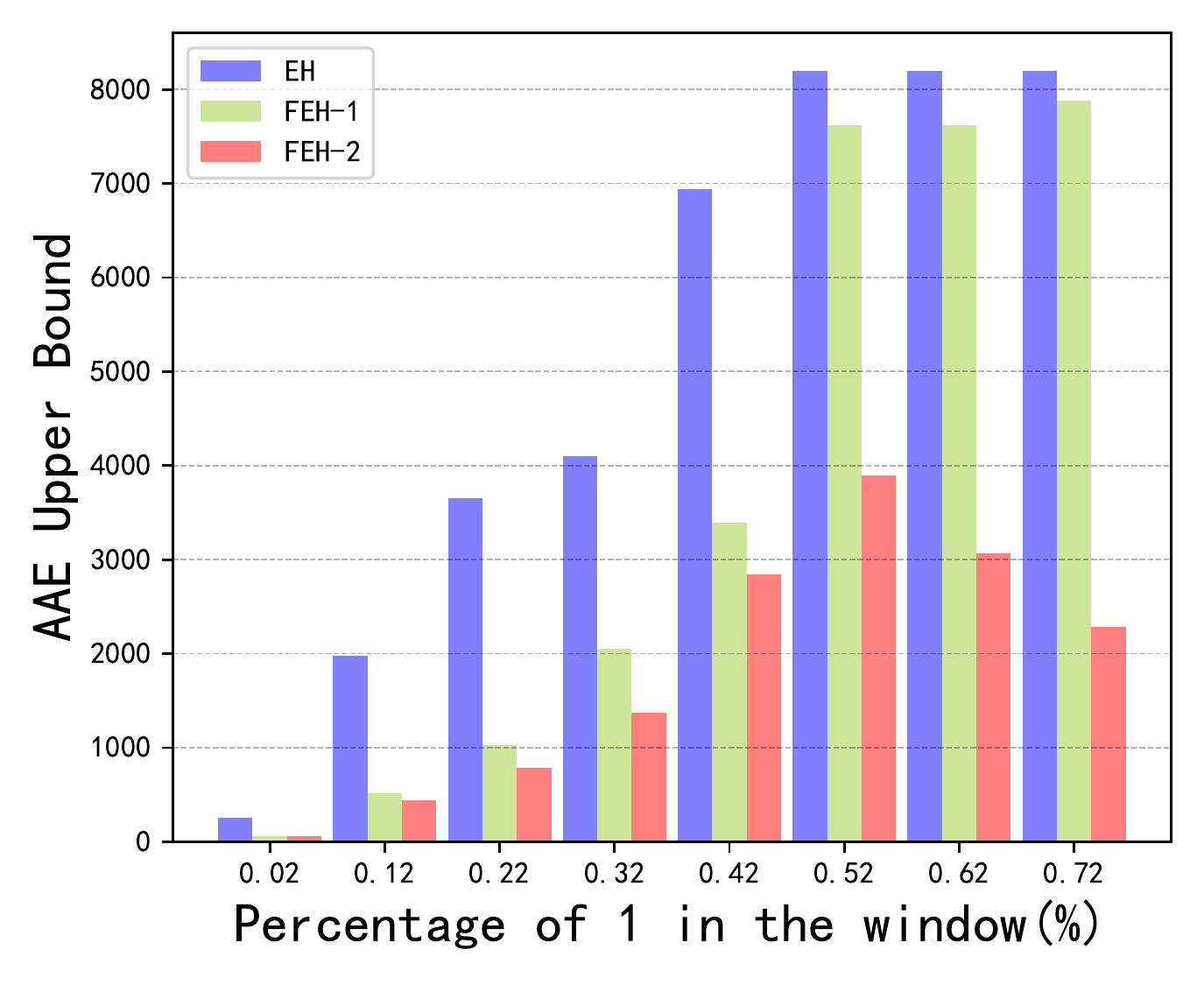}
		\label{fig10_3}}
	%	\hfil
	\subfloat[AREUB on campus dataset vs. percentage of 1 in the window]{\includegraphics[width=1.75in]{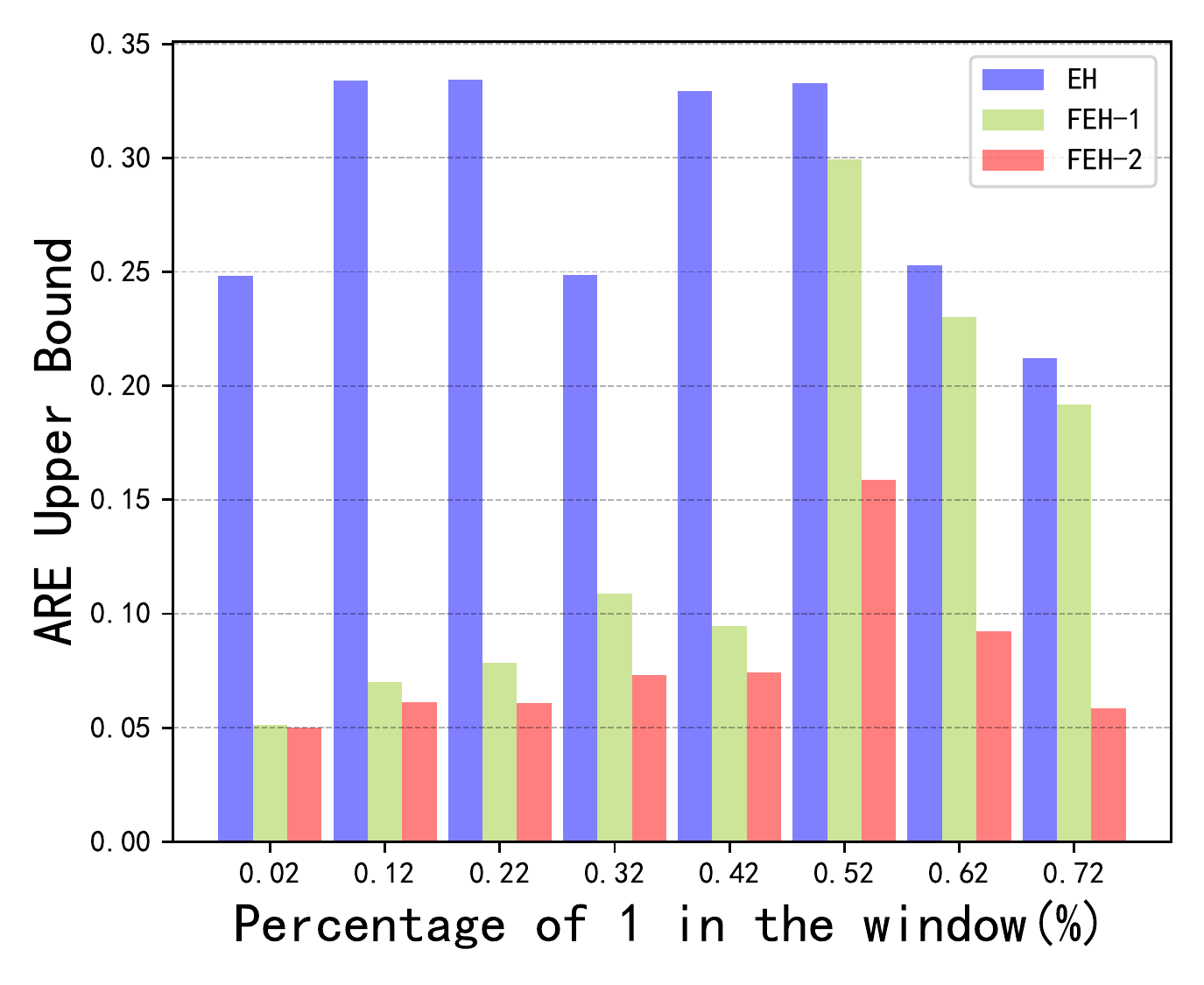}
		\label{fig10_4}}
	%	\hfil
	\caption{AAE and ARE of EH and P-FEH with different percentage of 1 in the window on Synthetic datasets}
	\label{fig10}
\end{figure*}

\subsection{The Speed of EH and P-FEH}
Fig.8 plots the update throughput and the query throughput of the EH and the P-FEH on different k increasing from 2 to 30 with a step of 4, and Fig.9 plots the update throughput and query throughput of the EH and the P-FEH on different N increasing from $2^{16}$ to $6^{*} 2^{16}$ with a step of $2^{16}$. The experimental results show that when facing various statistical requirements, the update and query throughput of P-FEH are basically the same as that of the EH. The word Acceleration technology we use in P-FEH reduces its extra cost of update. Therefore, when using the same memory, our P-FEH model can significantly improve the accuracy while the update and query speed remaining unchanged.

\subsection{Evaluation of P-FEH Query Strategy}
Fig.10 plots the AAEs, AREs, AAEUBs, AREUBs of the EH and the P-FEH on a different proportion of 1's in the window increasing from $2\%$ to $72\%$ with a step of $10\%$. The experimental results show that when the proportion of 1's in the window is larger, the flattened count of P-FEH improves the accuracy less, but the accuracy can still be greatly improved through the query strategy. This is because when the total count is large, there is a little empty bucket of EH that can be used by P-FEH to improve accuracy. However, according to the query strategy of P-FEH, in that time, the absolute error of the query results will be significantly reduced, and thus improving the overall accuracy. Therefore, our experimental results of accuracy have proved that the query strategy of P-FEH can further improve accuracy, and the accuracy will be improved more significantly when estimating elephant flows.

\section{CONCLUSION}
The solution of the Basic Counting problem can be used as building blocks to solve numerous more complex problems and has been applied to various fields. In this paper, we present a count-based sliding window model for this problem, the FEH model, which can achieve high accuracy and speed when using limited memory than the widely used EH mechanism. Experimental results demonstrate the performance improvements of our model. The FEH can be used to many complex problems in the field of data streams. We believe our paper can be a good help to the future study of the sliding window queries over data streams.

\end{document}